\documentstyle[preprint,aps]{revtex}
\tighten
\begin{document}
\preprint{THEF-NYM-93.02}
\hyphenation{Rijken}
\hyphenation{Nijmegen}
\hyphenation{Stich-ting Fun-da-men-teel On-der-zoek Ma-te-rie}
\hyphenation{Ne-der-land-se Or-ga-ni-sa-tie We-ten-schap-pe-lijk}

\title
{Antiproton-proton partial-wave analysis below 925 MeV/c}

\author{R. Timmermans,\cite{mail}
        Th.A. Rijken, and J.J. de Swart}

\address
{Institute for Theoretical Physics, University of Nijmegen,
 Nijmegen, The Netherlands\cite{email}}

\date{Accepted for publication in Phys. Rev. C }

\maketitle

\begin{abstract}
A partial-wave analysis of all $\overline{p}p$ scattering data
below 925 MeV/c antiproton laboratory momentum is presented.
The method used is adapted from the Nijmegen phase-shift
analyses of $pp$ and $np$ scattering data. We solve
the Schr\"odinger equation for the coupled $\overline{p}p$ and
$\overline{n}n$ channels where the long- and intermediate-range
interactions are described by a theoretically well-founded
potential. This gives the rapid variations of the scattering
amplitudes with energy. This potential consists of
the Coulomb potential with the main relativistic
correction, the magnetic-moment interaction,
the one-pion--exchange potential, and the heavy-boson
exchanges of the Nijmegen one-boson--exchange potential.
Slow variations of the amplitudes due to short-range
interactions, including the coupling to mesonic annihilation
channels, are parametrized by an energy-dependent, complex
boundary condition, specified at a radius of $r=1.3$ fm.
The Nijmegen 1993 $\overline{p}p$ database, consisting of
3646 $\overline{p}p$ scattering data, is presented
and discussed. The best fit to this database results
in $\chi^2_{\rm min}/N_{\rm data}$ = 1.043.
This good fit to the data shows that the Nijmegen long- and
intermediate-range potential is essentially correct.
The pseudovector coupling constant of the charged pion to nucleons
is determined to be $f^2_{\mbox{\scriptsize c}}$ = 0.0732(11)
at the pion pole, where the error is statistical.
\end{abstract}
\widetext

\section{INTRODUCTION}
A partial-wave analysis (PWA) of all antiproton-proton
($\overline{p}p$) scattering data below antiproton laboratory
momentum $p_{\mbox{\scriptsize lab}}=925$ MeV/c is presented.
In the field of nucleon-nucleon ($N\!N$) scattering,
phase-shift analyses (more properly called partial-wave
analyses, PWAs) have a long history and at present the multienergy or
energy-dependent partial-wave analyses of $N\!N$ scattering data
have reached a stage of considerable
sophistication~\cite{San83,Arn83,Arn87,Bys87,Ber88,Ber90,Klo93}.
Due to the poor quality
of low-energy antiproton beams and the resulting absence of
accurate experimental data, analogous model-independent
studies of the much more complex $\overline{p}p$ system have
in the past always been impossible.

In recent years, however, experimental progress has
been very significant, in particular due to the advent in 1983
of the LEAR facility at CERN. In the momentum region that we
consider in the analysis presented in this paper,
the situation between 400 and 925 MeV/c is quite good:
a variety of observables have been measured with
impressive accuracy. However, the practical difficulties
involved in constructing a high-quality antiproton beam of
even lower momentum are large. As a result, the $\overline{p}p$
database below about 400 MeV/c is still by far not as good as
one would like~\cite{Gra88,Mah88,Pir91}, in contrast
to the case of $N\!N$ scattering, where for instance very
accurate proton-proton differential cross sections have been
taken at energies as low as $T_{\mbox{\scriptsize lab}}=0.35$
MeV ($p_{\mbox{\scriptsize lab}}=25.6$ MeV/c).

During the last 10 years
a new method has been developed by the Nijmegen
group to perform partial-wave analyses of the $N\!N$
(proton-proton and neutron-proton) scattering data below
laboratory kinetic energy $T_{\mbox{\scriptsize lab}}=350$ MeV
($p_{\mbox{\scriptsize lab}}=883$ MeV/c).
The hallmark of this method is that the theoretical
knowledge of the $N\!N$ interaction is exploited
as much as possible in the description of the energy dependence
of the partial-wave scattering amplitudes. This is done by solving
the relativistic Schr\"odinger equation for each partial wave
and at each energy with the theoretically well-known long-range
$N\!N$ interaction represented by the potential for $r>b=1.4$ fm.
This long-range interaction is responsible for the fast
variation with energy of the scattering amplitudes.
The much slower energy variations  of the amplitudes
due to the short-range interactions are parametrized
phenomenologically by the energy dependence of a boundary
condition at $r=b$. In this way
one also avoids the complications due to the lack of
knowledge of this short-range interaction. In order to achieve a good
fit to the $N\!N$ data one has to include in the long-range
interaction the complete electromagnetic interaction (including
relativistic and some two-photon--exchange effects, the magnetic-moment
interactions, and the vacuum-polarization potential), the one-pion--exchange
potential, and also the intermediate-range forces (due to two-pion
exchange and/or heavier-meson exchange) of some realistic $N\!N$
potential model.

The dominant feature of $\overline{p}p$ scattering at low energies is the
annihilation into mesons, a process that has no counterpart in
$N\!N$ scattering. Annihilation is, in principle, a very complicated
spin-, isospin-, and energy-dependent multiparticle process.
At rest, where kinematically the production of 13 neutral pions
is allowed, 5 pions are produced on the average~\cite{Fet77}, and
of the order of 100 two-meson
channels contribute significantly~\cite{Van88,Loc92}.
In recent years some progress has
been made in understanding specific annihilation processes in
terms of quark-gluon degrees of freedom.
However, for the description of elastic $\overline{p}p \rightarrow
\overline{p}p$ and charge-exchange $\overline{p}p \rightarrow
\overline{n}n$ scattering, only a phenomenological
approach to annihilation is feasible at present.
In potential models for antinucleon-nucleon ($\overline{N}\!N$)
scattering one usually simplifies things drastically by taking
the annihilation potential to be completely independent
of spin, angular momentum, isospin, and energy.
This assumption is implemented either by applying a simple
absorptive boundary condition~\cite{Bal58,Spe67,Dal77,Del78},
or by using a state-independent two- or three-parameter
optical potential~\cite{Bry68,Myh77,Dov80,Koh86,Shi87,Hip89}.
However, when one is interested in describing the data
quantitatively, including spin-dependent observables,
a less naive approach is called for. For instance, in the
Paris $\overline{N}\!N$ model~\cite{Cot82,Pig91} a spin-, isospin-,
and energy-dependent optical potential is employed, and
in the Nijmegen coupled-channels model~\cite{Tim84}
each $\overline{N}\!N$ channel is coupled to two
effective two-body mesonic channels (for the coupled-channels
approach see also Ref.~\cite{Liu90}).
In both cases of the order of 15 parameters were needed to
obtain a more-or-less satisfactory fit to the pre-LEAR data, the
bulk of which consisted of elastic differential cross sections.
At that time charge-exchange data and spin-dependent observables were
practically absent. Because the $\overline{N}\!N$ system is so much more
complicated than the already quite complex $N\!N$ system it was
then believed that it would be impossible to perform a PWA
of the $\overline{p}p$ scattering data.

The complexity of the $\overline{p}p$ system when compared
to the $N\!N$ system (below the pion-production threshold)
is reflected in the following manner in a PWA.
In proton-proton ($pp$) scattering (isospin $I=1$) one has to specify
at each energy 2 phase shifts ($^1S_0$ and $^3P_0$) for
$J=0$ and on the average 2.5 phase parameters for each value
of the total angular momentum $J\neq 0$.
As an example: for $J=1$ one needs only one phase shift ($^3P_1$),
while for $J=2$ one needs 4 phase parameters
(the $^1D_2$, $^3P_2$, and $^3F_2$ phase shifts and the $\varepsilon_2$
mixing parameter). In a PWA of the
neutron-proton ($np$) scattering data (both isospin $I=0$ and $I=1$)
again 2 phase shifts are required for $J=0$, but now 5 phase parameters
are required for each value of $J\neq 0$.
Due to the lack of sufficient high-quality $np$ data
it has been impossible to do a good PWA of the $np$ data alone.
One needs to take the $I=1$ phases (with the exception of the $^1S_0$
phase shift) from the PWA of the $pp$ data and correct them for
electromagnetic and mass-difference effects ($M_p$ versus $M_n$ and
$m_{\pi^0}$ versus $m_{\pi^\pm}$).
In the case of $\overline{p}p$ scattering the Pauli
principle is not operative. Apart from this, the possibility
for the $\overline{p}p$ system to annihilate into mesons
complicates things even further.
This means that one has to determine no less than four times as
many phase parameters compared to the case of $np$ scattering,
so 8 phase parameters are required for $J=0$ and 20 phase
parameters for each value of $J\neq0$. In view of this,
the situation with regards to a $\overline{p}p$ PWA
indeed seemed quite hopeless.

Using essentially the same strategy as in the Nijmegen
multienergy partial-wave analyses of $N\!N$ scattering data,
and with the available recent high-quality data from LEAR and KEK,
we have nevertheless been able to perform a PWA
of $\overline{p}p$ scattering data below 925 MeV/c.
This work was started in 1987~\cite{Tim91c} and a preliminary report
of this analysis has already been given~\cite{Tim91b}.
The Schr\"odinger equation for the coupled
antiproton-proton and antineutron-neutron ($\overline{n}n$)
channels is solved. The short-range interaction,
including the coupling to the mesonic annihilation channels,
is parametrized by way of a complex boundary condition specified
at $r=b=1.3$ fm. The long-range interaction consists of the
Coulomb potential, the magnetic-moment interaction, and the
one-pion--exchange potential. The tail of the heavy-boson--exchange
part of the Nijmegen potential~\cite{Nag78} is used as
intermediate-range interaction. A lot of time and effort
has gone into collecting, scrutinizing, and
cleaning up the world set of $\overline{p}p$ scattering data,
which contains quite some flaws and contradictory data.
Exactly the same arguments were used in this process as were
used in the set-up of the Nijmegen $N\!N$
database~\cite{Ber88,Ber90,Klo93}. The resulting
Nijmegen 1993 $\overline{p}p$ database in the momentum interval
119--923 MeV/c consists of $N_{\rm data}=3646$ $\overline{p}p$
data, which can be fitted with $\chi^2_{\rm min}/N_{\rm data}=1.043$.
In view of the excellent quality of this
fit one can be confident that most of the amplitudes are
quite well described. The same methods have also been applied
by us to the strangeness-exchange reaction
$\overline{p}p \rightarrow \overline{\Lambda}\Lambda$,
which is an even more complicated process~\cite{Tim91a,Tim92a}.

This paper is organized as follows. In Sec.~II we present
the details of the method of analysis. In Sec.~III the
treatment of the short-range interaction by parametrizing
an energy-dependent boundary condition is discussed.
In Sec.~IV the long-range electromagnetic and pion-exchange
potential are presented and we specify the heavy-meson exchanges
used as intermediate-range interaction.
In Sec.~V we discuss the non-trivial problem how to
extract the nuclear scattering amplitude in the presence
of electromagnetic effects. The definition of the
phase-shift and mixing parameters for antinucleon-nucleon
scattering can be found in Sec.~VI. The statistical tools
used in the analysis are briefly reviewed in Sec.~VII. In
Sec.~VIII the Nijmegen 1993 antiproton-proton database is extensively
discussed. Next, in Sec.~IX we present the results of the analysis,
including the determination of the pion-nucleon coupling
constant. The most important results and conclusions are summarized
in Sec.~X. The algorithm to extract the phase parameters
from the $S$ matrix is reviewed in the Appendix.

\section{THE METHOD OF ANALYSIS}
The two-body scattering process is described with the coupled-channels
relativistic Schr\"odinger equation
\begin{equation}
     [ \Delta+p^2-2mV ] \: \psi({\bf r}) \: = \: 0  \:\: . \label{eq:Schrod}
\end{equation}
This is a matrix equation in channel space. We use the
physical particle basis $\overline{p}p$, $\overline{n}n$
in order to treat electromagnetic effects
properly and to account for the threshold of charge-exchange
scattering $\overline{p}p \rightarrow \overline{n}n$ at
$p_{\rm lab}=99.1$ MeV/c ($T_{\rm lab}=5.2$ MeV).
The connection between the channel momentum $p$ and the total energy
$\sqrt{s}$ in the center-of-mass system is given by the relativistic
expression $p^{2}=\frac{1}{4}s-m^{2}$ (for equal masses).
The relativistic Schr\"odinger equation Eqn.~(\ref{eq:Schrod})
is a differential form of the relativistic Lippmann-Schwinger integral
equation~\cite{Par70,Erk74}.
The difference between the relativistic and the ordinary
non-relativistic Lippmann-Schwinger equation~\cite{Lip50}
is the relation used between energy and momentum.
The relativistic Lippmann-Schwinger equation is
in turn equivalent to three-dimensional relativistic
integral equations like the Blankenbecler-Sugar
equation~\cite{Log63,Bla66,Kad68a,Kad68b}.
For a discussion about the derivation of potentials for use
in the relativistic Schr\"odinger equation, starting from the
field-theoretical Bethe-Salpeter equation~\cite{Sal51,Gel51}, see
for instance Refs.~\cite{Nag77,Swa78}.

The interaction in the region $r>b$ is described by
a theoretically well-founded antinucleon-nucleon potential.
This potential is given by
\begin{equation}
     V\: =\: V_C + V_{M\!M} + V_{N} \:\: ,
\end{equation}
where $V_C$ and $V_{M\!M}$ are the relativistic Coulomb and
magnetic-moment interaction respectively. $V_{N}$ is the
$\overline{N}\!N$ meson-exchange potential.
The precise forms of these potentials are discussed in Section IV.
After making the partial-wave projection by writing
\begin{equation}
   \psi({\bf r}) \: = \: \sum_{\ell s J m}
   \Phi^{m}_{\ell s J}(r)/r \:
   {\cal Y}^{m}_{\ell s J}(\theta) \:\: ,
\end{equation}
where
\begin{equation}
   {\cal Y}^{m}_{\ell s J}(\theta) \: = \: \sum_{m_\ell m_s}
   C^{\ell}_{m_\ell}\,^{s}_{m_s}\,^{J}_{m}\
   Y^{\ell}_{m_\ell}(\theta)\ \xi^{s}_{m} \:\: ,
\end{equation}
we obtain the radial Sch\"odinger equation, which
for partial waves with total angular momentum $J$ reads
\begin{equation}
  \left[ \frac{d^2}{dr^2} \: - \:
         \frac{L^2}{r^2}  \: + \: p^{2} \: - \: 2mV^{J}
         \right] \: \Phi^{J}(r) \: = \: 0  \:\: .
\end{equation}
$V^J$ is now a matrix with elements
$\langle \ell's'a'|V^{J}(r)|\ell\,s\,a\rangle$, where $a$
is an index used to distinguish the different channels.
For partial waves with $\ell=J$, $s=0,1$ all matrices are
$2\times2$, and for waves with $\ell=J\pm1$, coupled by the
tensor force, they are $4\times4$. The Schr\"odinger
equation is solved numerically~\cite{Ber87} starting with the
boundary condition at $r=b$ and ending at a value $r_{\infty}$
beyond the range of the nuclear potential.
At this point the $S$ matrix
is obtained by matching this numerical solution
$\Phi$ to the required asymptotic form
\begin{equation}
   \Phi_{\rm as}(r) \stackrel{r\rightarrow\infty}{\sim}
   \sqrt{\frac{m}{p}}
   \left[ H_{2}(pr)+H_{1}(pr) S^{J} \right]\:\: ,
   %\frac{1}{\sqrt{mp}} B \:\: .
\end{equation}
Since the Coulomb potential has infinite range one has to match to
Coulomb wave functions, so $H_1$ and $H_2$ are diagonal
matrices with entries $H^{(1)}_{\ell}(\eta_a,p_ar)$
and $H^{(2)}_{\ell}(\eta_a,p_ar)$,
the Coulomb analogues of the spherical Hankel functions.
$\eta_a$ is the relativistic Coulomb parameter
\begin{equation}
   \eta_a \: = \: \alpha/v_{\rm lab} \: = \: \alpha'\,m_a/p_a \:\: .
\end{equation}
Written in terms of the standard regular and irregular
Coulomb wave functions, $F_{\ell}(\eta,pr)$ and $G_{\ell}(\eta,pr)$,
these Hankel functions are defined as
\begin{equation}
   H^{(1)}_{\ell}(\eta_a,p_ar) = F_{\ell}(\eta_a,p_ar)-i
   G_{\ell}(\eta_a,p_ar)   \ \  , \ \
   H^{(2)}_{\ell}(\eta_a,p_ar) = F_{\ell}(\eta_a,p_ar)+i
   G_{\ell}(\eta_a,p_ar)   \:\: .
\end{equation}
These Coulomb wave functions read asymptotically
\begin{eqnarray}
   F_{\ell}(\eta_a,p_ar)
 & \stackrel{r\rightarrow\infty}{\sim} &
   \sin\left[p_ar-\ell\frac{\pi}{2}+
   \sigma_{\ell,a}-\eta_a\ln(2p_ar)\right] \:\: , \\
   G_{\ell}(\eta_a,p_ar)
 & \stackrel{r\rightarrow\infty}{\sim} &
   \cos\left[p_ar-\ell\frac{\pi}{2}+
   \sigma_{\ell,a}-\eta_a\ln(2p_ar)\right] \:\: ,
\end{eqnarray}
where the Coulomb phase shifts $\sigma_{\ell,a}$ are given by
\begin{equation}
   \sigma_{\ell,a} \: = \: \arg\,\Gamma(\ell+1+i\eta_a)
   \:\: .
\end{equation}
In case the Coulomb interaction is absent in a
channel (like $\overline{n}n$), one can simply put $\eta_a=0$.
Then the Coulomb wave functions become
\begin{eqnarray}
   F_{\ell}(0,\rho) = \rho j_{\ell}(\rho) \:\:\: & , & \:\:\:
   G_{\ell}(0,\rho) = -\rho n_{\ell}(\rho) \:\: , \nonumber \\
   H^{(1)}_{\ell}(0,\rho) = \rho h^{(1)}_{\ell}(\rho) \:\:\: & , & \:\:\:
   H^{(2)}_{\ell}(0,\rho) = \rho h^{(2)}_{\ell}(\rho) \:\: ,
\end{eqnarray}
in terms of ordinary spherical Bessel, Neumann, and Hankel functions.
The matching procedure at $r_{\infty}$ works as follows.
The multichannel Wronskian is defined by
\begin{equation}
   W(\Phi_{1},\Phi_{2}) \: = \:
   \Phi_{1}^{\rm T} \frac{\textstyle 1}{\textstyle m} \Phi'_{2} -
   \Phi_{1}^{\prime\:{\rm T}} \frac{\textstyle 1}{\textstyle m} \Phi_{2}
   \:\: ,
\end{equation}
where the prime denotes differentiation with respect to $r$
and the ``T'' denotes transposition in channel space.
The Wronskian of two arbitrary solutions $\Phi_1$ and $\Phi_2$ is
independent of $r$ and equal to 0 because of the boundary condition
$\Phi(0)=0$.
Demanding that
\begin{equation}
   W(\Phi(r_{\infty}),\Phi_{\rm as}(r_{\infty}))
   \: \equiv \: 0 \:\: ,
\end{equation}
we obtain for the partial-wave $S$ matrix
\begin{equation}
   S^{J} \: = \: - \left[ \Phi^{\prime{\rm T}}
   \frac{\textstyle 1}{\textstyle \sqrt{mp}} H_{1} - \Phi^{\rm T}
   \sqrt{\frac{\textstyle p}{\textstyle m}} H'_{1} \right]^{-1}
   \left[ \Phi^{\prime{\rm T}}
   \frac{\textstyle 1}{\textstyle \sqrt{mp}} H_{2} - \Phi^{\rm T}
   \sqrt{\frac{\textstyle p}{\textstyle m}} H'_{2} \right] \:\: ,
   \label{eq:smtrx}
\end{equation}
where the prime on the Hankel functions denotes differentiation
with respect to the argument $pr$.
Since the matching is to Coulomb wave functions, what is actually
obtained in this manner is the partial-wave $S$ matrix with respect
to the Coulomb force, denoted by $S^C_{C+M\!M+N}$. How
to calculate the scattering amplitude
is explained in Section V. The final step, the calculation of
all the observables from the scattering amplitude, is
standard~\cite{Wol52,Hos68,Bys78,LaF80,Bys84}. See in particular
Ref.~\cite{LaF92} for the case of antinucleon-nucleon scattering.

\section{THE SHORT-RANGE INTERACTION}
The short-range dynamics is treated
with the help of a boundary condition applied at
a distance $r=b$.
This boundary condition, called the $P$~matrix,
is the logarithmic derivative of the solution
matrix $\Phi(r)$ at a distance $r=b$ from the origin
\begin{equation}
   P \: = \: b \left(\frac{d\Phi}{dr}\Phi^{-1}\right)_{r=b} \:\: .
\end{equation}
The factor $b$ is included in this definition in order to
make the $P$ matrix dimensionless. The boundary-condition
approach to strong interactions goes back to the work of Feshbach
and Lomon~\cite{Fes64} and earlier. The term $P$ matrix was introduced
by Jaffe and Low~\cite{Jaf79,Jaf83} for use with the bag model where at
the energies of the eigenstates of the confined quark and gluon
degrees of freedom the $P$ matrix exhibits poles that are
not necessarily also present in the $S$ matrix. (For a review,
see Ref.~\cite{Bak86}.) At this stage,
we will not attempt to make any connection with multiquark states.
In general, this boundary  relates the inner- to the outer-region
physics. In $N\!N$ and $\overline{N}\!N$ scattering the
short-range interaction
is essentially unknown and has to be treated phenomenologically.
The long-range physics one understands theoretically much better.
The $P$-matrix formalism provides a useful separation between
these two regions and has become a powerful tool in analyzing scattering
data. In the Nijmegen partial-wave analyses of $N\!N$
scattering data~\cite{Ber88,Ber90,Klo93} the $P$-matrix method has
already proven its power~\cite{Swa85}.

For the parametrization of the partial-wave $P$ matrix, it is
convenient to use a very simple model for the short-range
dynamics. We assume that the interaction in each partial wave
can be described by a spherical well, a potential which may
depend on spin, isospin, and energy, but which is constant
as a function of distance. The $P$ matrix for such a potential
can be evaluated analytically. For a single-channel spherical-well
problem in a partial wave with orbital angular momentum $\ell$
it is given by
\begin{equation}
  P_{\ell} \:=\: p'b \: J_{\ell}'(p'b)/J_{\ell}(p'b) \:\: ,
\end{equation}
where $J_{\ell}(\rho)=\rho j_{\ell}(\rho)$ with $j_{\ell}(\rho)$
the spherical Bessel function and ${p'}^{2}=p^{2}-2mV$, $V$ being
the depth of the spherical-well potential. The prime on the
Bessel function denotes differentiation with respect to the
argument. In the case of $\overline{N}\!N$ scattering
we take these short-range potentials to be complex in order
to account for the annihilation into mesonic channels. The
short-range interaction is in this manner described by a simple
state-dependent optical potential~\cite{Fes58}.
It turns out that we can take the short-range spherical-well potential
independent of the energy and still fit the data properly.
This was not possible in the Nijmegen partial-wave analyses of the
$N\!N$ data: there we had to make the short-range potential energy
dependent. That this is not necessary for $\overline{N}\!N$ scattering
is probably because of the absence of high-quality data
at low energies such as present in $N\!N$ scattering.
It is, however, crucial that we take the real parts to be dependent
on spin and isospin. It also turned out that the imaginary parts
of these short-range potentials could be taken independent
of isospin. Each partial wave is thus parametrized
by a maximum of 3 parameters, a complex spherical well
for both isospin 0 and 1, where the imaginary parts are equal
for both isospins.  How many and which parameters
are actually needed in each individual partial wave is
discussed below in Section IX.

In the Nijmegen analyses of $N\!N$ scattering data,
a value of $b=1.4$ fm for the boundary radius was found
to be suitable. In the $\overline{N}\!N$ case, the results are
rather sensitive on the choice of the value of $b$.
The best results are obtained with $b=1.3$ fm.
Since we use in the outer region a real potential consisting
of an electromagnetic and a meson-exchange part, the coupling
to the mesonic channels is completely absorbed in the
boundary condition. The radius $b$ is therefore a clear
measure for the range of the annihilation potential.
The fact that the results are quite sensitive on $b$
shows that this range is in fact approximately 1.3 fm.
So we find definite indication that annihilation in
$\overline{p}p$ scattering is a rather long-range process.

It remains to discuss the parametrization of the $P$~matrix
for the states with $\ell=J\pm1$ coupled by the tensor force.
In these cases it is convenient to use the method also
employed in the $N\!N$ case. For a certain value
of the isospin, we start with a diagonal $2\times2$ $P$~matrix
and use additional parameters to describe the short-range
mixing between the two coupled partial waves, as follows
\begin{equation}
  P \:\: = \:\: \left( \begin{array}{rr}
                        \cos\theta  & \sin\theta    \\
                       -\sin\theta  & \cos\theta \end{array} \right)
                   \left( \begin{array}{cc}
                           P_{1} & 0       \\
                             0   & P_{2}   \end{array} \right)
                 \left( \begin{array}{rr}
                       \cos\theta  & -\sin\theta    \\
                       \sin\theta  &  \cos\theta \end{array} \right)
                                \:\: .
  \label{eq:mix}
\end{equation}
The mixing angle $\theta$ can, if necessary, be parametrized
as a function of energy, but again in our analysis we can
take it in all cases independent of the energy. We need these
mixing angles only for the isospin $I=0$ states, for the $I=1$
states they can be set to zero.

Time-reversal invariance allows us to choose the phases
of the physical states in such a way that the potential
matrix is symmetric. The full $S$ matrix, including all
mesonic channels, is then also symmetric and of course unitary.
When we restrict ourselves to the $\overline{N}\!N$ channels,
then this sub-block of the $S$ matrix is of course still
symmetric, but no longer unitary. This reflects the
disappearance of probability (flux) into the mesonic channels.
Correspondingly, the $P$ matrix in our case is still symmetric,
but not hermitian, as in $N\!N$ scattering (below
the pion-production threshold).

Why is the $P$-matrix approach to partial-wave analyses
so convenient and powerful? One of the main reasons is that
it allows an easy parametrization of the energy dependence
of the scattering amplitudes. This can be seen as follows.
The long-range interactions lead to rapid variations
with energy of the amplitudes. These variations are
much easier to parametrize when one uses the $P$ matrix
than the $S$ matrix (or $K$ matrix).
For instance, in the presence of the Coulomb interaction,
the $S$ matrix has an essential singularity and a branchpoint
at zero energy. However, if the Coulomb potential is included
in the potential tail, these singularities and
the corresponding left-hand cut are absent from the
$P$ matrix. Similarly, the left-hand cuts due to all meson exchanges
included in the potential tail are absent from the $P$ matrix.
All these cuts, however, are present in the $S$ matrix, in
addition to the kinematical unitarity cut. The point is that
the slow variations with energy of the amplitudes due to short-range
interactions are easy to parametrize, once the rapid variations
have been taken care of by explicitly including the corresponding
long-range potentials in the Schr\"odinger equation.
Of course, some left-hand cuts remain, such as the cut due to
uncorrelated two-pion exchange, as well as right-hand cuts due
to the coupling to inelastic channels. These, however, lead to
much slower energy variations of the amplitudes than the long-range
electromagnetic interactions and one-pion exchange.

\section{THE POTENTIAL TAIL}
In order to obtain a good description of the data, the
long-range interaction between the particles, consisting of
the electromagnetic interaction and the one-pion--exchange potential,
must be included properly. These interactions are model independent
in the sense that they are (or at least should be) the same in all
models of the (anti)nucleon-nucleon force. The fit to the data
is improved when realistic meson-exchange forces are used as
intermediate-range potential.

The long-range electromagnetic potential is included
to order $\alpha=e^2/4\pi$, the fine-structure constant.
The one-photon--exchange potential is derived from the
phenomenological electromagnetic Lagrangian
\begin{equation}
  {\cal L}_{\gamma} =  e\:Q\:[i\overline{\psi}\gamma_{\mu}\psi]\:
  A^{\mu} \: + \: e\:\frac{\kappa}{4M}\:[\overline{\psi}
  \sigma_{\mu\nu}\psi]\: (\partial^{\mu}A^{\nu}-
  \partial^{\nu}A^{\mu}) \:\: ,
\end{equation}
where $Q$ is the nucleon charge in units of $e>0$ and $\kappa$ %$\kappa=\mu-1$
is the anomalous magnetic moment, for proton and neutron
$\mu_{p} = 1+\kappa_{p} = 2.793$, and
$\mu_{n} = \kappa_{n} = -1.913$, respectively. $\psi$
is the proton or neutron field and $A^{\mu}$ the photon field.
If the spatial extension of the nucleons is taken into account,
the charge and magnetic moments are in momentum space replaced by the
Dirac form factor $F_1(t)$ and the Pauli form factor $F_2(t)$,
which are functions of the four-momentum transfer $t$.
The static limits $t=0$ are
\begin{equation}
F^{p}_{1}(0) = 1 \:\: , \:\: F^{n}_{1}(0) = 0 \:\: , \:\:
F^{p}_{2}(0) = \frac{ \kappa_{p}}{ 2M_{p}} \:\: , \:\:
F^{n}_{2}(0) = \frac{ \kappa_{n}}{ 2M_{n}} \:\: .
\end{equation}
In the point-particle approximation the momentum dependence of these
form factors, reflecting the inner structure of the nucleons,
is neglected. In this approximation we end up with the
spin-dependent one-photon--exchange potentials for $r>b$
\begin{equation}
   V_{\gamma}(r) \: = \: -\frac{ \alpha'}{ r} \: + \:
   \frac{\mu^{2}_{p}}{4M^{2}_{p}}\:\:
   \frac{\alpha}{ r^{3}}\:S_{12}
   \: + \: \frac{8\mu_{p}-2}{ 4M^{2}_{p}}
   \frac{\alpha}{ r^{3}}\:{\bf L}\cdot{\bf S} \:\:\:\:\:
   {\rm for} \:\: \overline{p}p \rightarrow \overline{p}p \:\: ,
   \label{eq:Vmm1}
\end{equation}
and
\begin{equation}
   V_{\gamma}(r) \: = \: \frac{\mu^{2}_{n}}{4M^{2}_{n}}
   \frac{\alpha}{ r^{3}}\:S_{12} \:\:\:\:\:
   {\rm for} \:\: \overline{n}n \rightarrow \overline{n}n \:\: .
   \label{eq:Vmm2}
\end{equation}
These potentials are obtained by calculating the one-photon--exchange
diagrams in momentum space and applying a Fourier transformation
to configuration space. The momentum dependence of the form
factors can be taken into account. This leads to
short-range modifications of the one-photon--exchange potential.
Since we use only the potential outside $r=b=1.3$ fm,
we did not include these effects. The use
of $\alpha'$ in the central potential for
$\overline{p}p \rightarrow \overline{p}p$ takes care
of the main relativistic corrections to the Coulomb
potential~\cite{Bre55,Aus83}.
It is given by
\begin{equation}
   \alpha' \: = \: \alpha \: 2p/(Mv_{\rm lab}) \:\: ,
\end{equation}
where $v_{\rm lab}$ is the velocity of the
antiproton in the laboratory system.
In order to appreciate the order of magnitude of this factor,
at 600 MeV/c for instance, where $v_{\rm lab}=0.54$,
one has $\alpha'=1.135\ \alpha$.
The spin-orbit potential comes from the interaction of the magnetic
moment of one particle with the Coulomb field of the other particle
(and is consequently absent in
$\overline{n}n \rightarrow \overline{n}n$). It includes
a relativistic correction due to the Thomas precession.
The tensor potential comes from the interaction of the two
magnetic moments. In our energy range the Coulomb and the
magnetic-moment interaction are the dominant electromagnetic
effects. The vacuum-polarization potential~\cite{Dur57}, which is
important~\cite{Ber88} in low-energy $pp$ scattering
(below $T_{\mbox{\scriptsize lab}}=$ 30 MeV),
has a negligible influence here. Two-photon--exchange
effects~\cite{Aus83} are neglected as well.

The following simple one-pion--exchange potential without
a form factor is used
\begin{equation}
 V_{\pi}(r) \: = \: f^2_{N\!N\pi} \frac{M}{\sqrt{p^2+M^2}}
      \frac{m^2}{m^2_{\pi^{\pm}}} \frac{1}{3}
         \left[ \mbox{\boldmath $\sigma$}_{1}\cdot
               \mbox{\boldmath $\sigma$}_{2} + S_{12}
            \left(1+\frac{3}{(mr)}+\frac{3}{(mr)^{2}} \right)\right]
            \frac{e^{-mr}}{r}  \;\; .
\end{equation}
The mass difference between the neutral $\pi^0$
and charged $\pi^{\pm}$ pion is included, so we take
$m=m_{\pi^0}$ for the elastic reactions
$\overline{p}p \rightarrow \overline{p}p$ and
$\overline{n}n \rightarrow \overline{n}n$, and
$m=m_{\pi^{\pm}}$ for the off-diagonal charge-exchange transitions
$\overline{p}p \leftrightarrow \overline{n}n$.
In principle, the pion-nucleon coupling constant is charge
dependent due to the mass difference between the up and down quarks
and due to the electromagnetic corrections. We introduce here the
relevant coupling constants $f_{N\!N\pi}^2$ at the different
vertices
\begin{equation}
 f^{2}_{\mbox{\scriptsize p}} \equiv
          f^{2}_{\mbox{\scriptsize pp$\pi^{0}$}} \:\: , \:\:
 f^{2}_{\mbox{\scriptsize n}} \equiv
          f^{2}_{\mbox{\scriptsize nn$\pi^{0}$}} \:\: , \:\:
 2f^{2}_{\mbox{\scriptsize c}} \equiv
  f_{\mbox{\scriptsize pn$\pi^{+}$}} f_{\mbox{\scriptsize np$\pi^{-}$}}\:\: ,
\end{equation}
for the reactions $\overline{p}p \rightarrow \overline{p}p$,
$\overline{n}n \rightarrow \overline{n}n$, and
$\overline{p}p \leftrightarrow \overline{n}n$ respectively.
In the PWA a charge-independent pion-nucleon coupling constant is used,
where $f^2_{\mbox{\scriptsize p}} = f^2_{\mbox{\scriptsize n}} =
 f^2_{\mbox{\scriptsize c}}$ = 0.0745,
taken from the most recent Nijmegen $pp$ PWA~\cite{Sto93}.
In Sect. IX, however, this assumption is relaxed when we
determine the charged-pion coupling constant
from the charge-exchange data by adding $f^2_{\mbox{\scriptsize c}}$
as a free parameter.

To improve the description of the data,
some sort of heavy-meson--exchange potential has to be added to
one-pion exchange in the outer region. Since the tails of realistic
$N\!N$ potentials are remarkably similar, it probably
does not matter very much which potential one picks, provided
it gives a good description of the $N\!N$
scattering data. We have opted for the
charge-conjugation--transformed version of the Nijmegen
one-boson--exchange soft-core $N\!N$ potential~\cite{Nag78},
which is one of the best $N\!N$ potentials available.
(The use of $C$ conjugation rather than $G$ conjugation is
more natural when one works on the physical particle basis.)
The tail of this potential has already been used in
the Nijmegen $pp$ and $np$ partial-wave analyses, and the tail
of the corresponding Nijmegen soft-core hyperon-nucleon
potential~\cite{Mae89,Swa89} has been used by us in our PWA
of the reaction $\overline{p}p
\rightarrow \overline{\Lambda}\Lambda$~\cite{Tim91a,Tim92a}.
The following heavy-boson--exchange potentials are included.

$\bullet$ Pseudoscalar-meson exchange.
Included are $\eta(549)$ and $\eta'(958)$ exchange.
For the pseudoscalars, including the pion, we use
the pseudovector type of Lagrangian
\begin{equation}
    {\cal L}_{\mbox{\scriptsize PV}} =
       \sqrt{4\pi}\frac{f}{m_{\pi^{\pm}}}\:[i\overline{\psi}
       \gamma_{\mu}\gamma_{5}\psi]\:
       \partial^{\mu}\Phi_{\mbox{\scriptsize P}}  \:\: .
\end{equation}
Although equivalent to the pseudoscalar type of interaction for
one-meson exchange between protons, the pseudovector interaction
is favored because it gives a more reasonable
two-pion--exchange potential and because it leads to
at most small breakings of flavor symmetry for the coupling
constants of the pseudoscalar nonet to baryons~\cite{Tim91a}.
The scaling mass $m_{\pi^{\pm}}$ is conventionally introduced
to make the coupling constant $f$ dimensionless.

$\bullet$ Vector-meson exchange.
Included are $\rho(770)$, $\omega(782)$, and $\phi(1019)$ exchange.
The Lagrangian (in terms of rationalized coupling constants) is
\begin{equation}
    {\cal L}_{\mbox{\scriptsize V}} =
       \sqrt{4\pi}g\:[i\overline{\psi}\gamma_{\mu}\psi]\:
       \Phi^{\mu}_{\mbox{\scriptsize V}}
       + \sqrt{4\pi}\frac{f}{4M_p}\:[\overline{\psi}
       \sigma_{\mu\nu}\psi]\:
       (\partial^{\mu}\Phi^{\nu}_{\mbox{\scriptsize V}}-
       \partial^{\nu}\Phi^{\mu}_{\mbox{\scriptsize V}}) \:\: .
\end{equation}

$\bullet$ Scalar-meson exchange. We include $a_0$(783),
$f_0$(975), and $f'_0$(760) exchange. The Lagrangian is
\begin{equation}
    {\cal L}_{\mbox{\scriptsize S}} =
       \sqrt{4\pi}g\:[\overline{\psi}\psi]\:
       \Phi_{\mbox{\scriptsize S}} \:\: .
\end{equation}
The scalar mesons have always been a controversial topic.
In early one-boson--exchange models for the $N\!N$ interaction
there was a clear need for an isoscalar scalar ``$\sigma$'' meson
with an effective mass of about 550 MeV~\cite{Hos60,Hos62,Bry64}.
While no such low-mass particle exists, there was some evidence
in production experiments for a broad structure
$\varepsilon$(760) under the $\rho^0$, often explained
away as a strong $\pi\pi$ final-state interaction. Later
it was pointed out~\cite{Bin71,Bry72,Swa78} that such a wide
($\Gamma \approx$ 640 MeV)
$\varepsilon$(760) simulates the narrow-``$\sigma$'' exchange
in one-boson--exchange models for the $N\!N$ interaction.

The situation in phase-shift analyses of $\pi\pi$ scattering
data, obtained from reactions as $\pi N \rightarrow \pi\pi N$,
has for a long time been confusing and not conclusive~\cite{Pro73}.
In these analyses the assumption has always been that only
$\pi$ exchange is relevant, while $a_1$ exchange can be neglected.
Very recently, however, the situation
has been much clarified~\cite{Sve92b}. Data
on $\pi N\!\uparrow \rightarrow \pi^+\pi^- N$ using a polarized
target provide unambiguous evidence for a broad $I=0$ $0^{++}$(750)
state, when a proper amplitude analysis is done, including also
$a_1$ exchange. In a similar amplitude analysis
of data on $K^+ n\!\uparrow \rightarrow
K^+ \pi^- p$~\cite{Sve92a} evidence is found for
$I=1/2$ $0^+$(887) strange scalar mesons under the $K^*$(892).

In the quark model, several mechanisms
give rise to scalar ($J^P=0^+$) mesons. The simplest
model is the $^3P_0$ $\overline{Q}Q$ states. Then there
are the glueball states and the cryptoexotic $\overline{Q}^2Q^2$
states~\cite{Jaf77}. A physical scalar meson
will in general be a mixture of
$\overline{Q}Q$, $\overline{Q}^2Q^2$, and glueball components.
The $\overline{Q}Q$ states are expected~\cite{Aer80} near the
other $^3P$ $\overline{Q}Q$ mesons, that is around 1250 MeV.
Glueballs are also not very likely to exist below 1000
MeV~\cite{Sha84}. For the $\overline{Q}^2Q^2$
states, however, one~\cite{Jaf77} does predict a low-lying nonet
of scalar mesons. The lowest state, with only nonstrange
quarks, has $I=0$ and decays into $\pi\pi$. It can be
identified with the $\varepsilon$(760) under the $\rho^0$.
This nonet contains also
a nearly degenerate set of $I=0$ and $I=1$ cryptoexotic scalar
mesons (like the $\rho$(770) and $\omega$(782)) with
an $\overline{s}s$ pair. These are easily identified
as the $f_0(975)$ and $a_0$(983) mesons, previously
called $S^*$(975) and $\delta$(983) respectively, with
their relatively large branching ratios into $\overline{K}K$.
The nonet is completed by a set of broad $I=1/2$ strange mesons
$K^*_0$(887) seen~\cite{Eng78} under the $K^*$(892).

$\bullet$ Next to these conventional mesons, the Nijmegen
soft-core potential (originally derived from Regge-pole theory) also
contains pomeron exchange, which in QCD is understood as
color-singlet two- or multigluon exchange~\cite{Low75,Nus75,Sim90},
and the weak diffractive scalar-like part of the tensor-meson
exchanges. The short-range non-local terms of the potential
are neglected, which is a very good approximation
outside $r=b=1.3$ fm.

The consequences of meson exchanges
for the $\overline{N}\!N$ interaction
have been examined by many authors, for instance by
Dover and Richard~\cite{Dov78,Dov79,Buc79}.
It turns out from a qualitative investigation
that while in the $N\!N$ case there is a
strong coherence between the isospin $I=1$ spin-orbit forces, in the
$\overline{N}\!N$ case very strong tensor forces occur, especially
for isospin $I=0$. In the $N\!N$ system the
vector $\omega$(782) and the scalar $f'_{0}$(760) exchange
make up the strong spin-orbit force that splits the $^3P_{0,1,2}$
phase shifts, but the central potentials of these exchanges
largely cancel each other. Similarly, the spin-orbit forces
due to the exchange of the vector $\rho$(770) and the
scalar $a_{0}$(983) add up, but the central potentials cancel.
Applying charge conjugation to the different meson exchanges,
one sees that in the $\overline{N}\!N$ potential the central
potentials from $\omega$(782) and $f'_{0}$(760) exchange
add coherently to a very strong attractive potential.
This has led to speculation about the existence of
$\overline{N}\!N$ bound states and
resonances~\cite{Sch71,Bog74,Sha78,Mon80,Car91}.
The tensor potentials from $\rho(770)$ and $\pi(140)$ exchange
also add up and dominate the charge-exchange
reaction $\overline{p}p \rightarrow \overline{n}n$.
The same phenomenon is present in the reaction
$\overline{p}p \rightarrow \overline{\Lambda}\Lambda$,
where $K$(494) and $K^*$(892) exchange conspire to build up
the strong tensor force that is the hallmark of this
reaction~\cite{Tim91a}.

\section{THE AMPLITUDES}
The evaluation of the scattering amplitude in
the presence of electromagnetic effects is a
non-trivial problem that requires special care.
In the $pp$ PWA the precise values and
energy dependence of the phase shifts are influenced
significantly by the electromagnetic interaction.
The same will obviously also be true in the $\overline{p}p$ case.

We start by discussing the case where next to the nuclear
force we only have the Coulomb potential to deal with.
The scattering amplitude due to the Coulomb force is given by
\begin{eqnarray}
   \langle s'm'a'|M_C(\theta)| s\,m\,a\rangle & = &
   -\delta_{ss'}\delta_{mm'}\delta_{aa'} \,
   \frac{\eta_a}{p_a(1-\cos\theta)}
   e^{-i\eta_a\ln\frac{1}{2}(1-\cos\theta)+2i\sigma_{0,a}}
   \nonumber \\ & = &
   -\delta_{ss'}\delta_{mm'}\delta_{aa'} \, \frac{\eta_a}{2p_a}
   \frac{e^{2i\sigma_{0,a}}}{(\sin^2\frac{1}{2}\theta)^{1+i\eta_a}}
   \:\: . \label{eq:coul}
\end{eqnarray}
The partial-wave decomposition of $M_C(\theta)$ in terms of
Coulomb phase shifts and Legendre polynomials does not converge
point-like, due to the infinite range of the Coulomb
potential. However, it can be summed in the sense of
distributions~\cite{Tay74,Sem75,Ger76} to give the Coulomb
amplitude Eqn.~(\ref{eq:coul}). In order to make a partial-wave
decomposition of the scattering amplitude, the
total scattering amplitude $M_{C+N}(\theta)$ is split as follows
\begin{equation}
   M_{C+N}(\theta) \: = \: M_C(\theta) + M^C_{C+N}(\theta) \:\: ,
\end{equation}
where the amplitude $M^C_{C+N}(\theta)$ is the nuclear
scattering amplitude in the presence of the Coulomb potential.
Its partial-wave decomposition reads
\begin{eqnarray}
   \langle s'm'a'| M^C_{C+N}(\theta) | s\,m\,a \rangle & = &
   \sum_{\ell\,\ell' J} \sqrt{4\pi(2\ell+1)} \: i^{\ell-\ell'} \:
   C^{\ell}_{0}\,^{s}_{m}\,^{J}_{m} \:
   C^{\ell'}_{m-m'}\,^{s'}_{m'}\,^{J}_{m} \:
   Y^{\ell'}_{m-m'}(\theta) \nonumber \\
  & & \langle\ell's'a'|S_{C}^{1/2}(S^{C}_{C+N}-1)S_{C}^{1/2}|\ell\,s\,a\rangle
   / 2ip_a \:\: , \label{eq:ampl}
\end{eqnarray}
where $S^{C}_{C+N}$ is the nuclear $S$ matrix in the presence of
the Coulomb force.  The total $S$ matrix $S_{C+N}$, due to the Coulomb
and nuclear interaction, is then given by
%multiplied by the Coulomb $S$ matrix, as follows
\begin{equation}
   S_{C+N} \: = \: S_C^{1/2} \, S^C_{C+N} \, S_C^{1/2} \:\: .
\end{equation}
$S_C$ is the Coulomb $S$ matrix with matrix elements
\begin{equation}
 \langle \ell's'a'|S_C|\ell\,s\,a\rangle \: = \:
 \delta_{\ell\ell'}\delta_{ss'}\delta_{aa'} \,
 \exp(2i\sigma_{\ell,a}) \:\: .
\end{equation}
The nuclear $S$ matrix in the presence of the Coulomb force
$S^C_{C+N}$ is obtained by solving the Schr\"odinger equation
numerically and matching to Coulomb wave functions
(see Section II). It is given by Eqn.~(\ref{eq:smtrx}).

Next we discuss the generalization
to the case where next to the Coulomb force also the
magnetic-moment interaction is present~\cite{Sto90}. Although
this latter potential has a finite range and consequently the
partial-wave decomposition converges, it is still much more practical
to split off in the scattering amplitude the contribution
of the magnetic-moment interaction. In this way
the magnetic-moment interaction can be included in all partial
waves and the summation in the nuclear amplitude converges much faster.
We thus write
\begin{equation}
  M_{C+M\!M+N}(\theta) \: = \:  M_C(\theta) + M^C_{C+M\!M}(\theta)
                          + M^{C+M\!M}_{C+M\!M+N}(\theta) \:\: .
\end{equation}
Here $M^C_{C+M\!M}(\theta)$ is the scattering amplitude of the
magnetic-moment interaction in the presence of the Coulomb force.
Since this amplitude is almost exactly in phase with
the Coulomb scattering amplitude $M_C(\theta)$, it is essential
that the effect of the magnetic-moment interaction
is evaluated in Coulomb-distorted-wave Born approximation (CDWBA),
and not in plane-wave Born approximation, as was pointed out
by Knutson and Chiang~\cite{Knu78}. The contribution, in CDWBA,
of the magnetic-moment interaction to the $K$ matrix is
\begin{equation}
 \langle \ell's'a'|K_{M\!M}|\ell\,s\,a\rangle \: = \:
   -\delta_{aa'}\frac{M_a}{p_a}\int^{\infty}_{0}\!dr\:
 F_{\ell'}(\eta_{a'},p_{a'}r) V_{M\!M}(r) F_{\ell}(\eta_{a},p_{a}r)
   \:\: .
\end{equation}
Integrals of this type can be evaluated very rapidly and
accurately by a backward-recursion algorithm~\cite{San99}.
The magnetic-moment interaction $V_{M\!M}$ is given in
Eqn.~(\ref{eq:Vmm1}) for $\overline{p}p \rightarrow \overline{p}p$
and in Eqn.~(\ref{eq:Vmm2}) for $\overline{n}n \rightarrow \overline{n}n$.
All these partial-wave contributions are subsequently summed
to obtain the total amplitude $M^C_{C+M\!M}(\theta)$.
In practice it turns out that the spin-orbit potential
of the magnetic-moment interaction leads to a contribution
$Z_{LS}$ to $\langle11a|M^C_{C+N}(\theta)|10a\rangle$
that converges much too slowly to be summed term by term.
This part can be summed analytically~\cite{Knu78}.
For antiproton-proton scattering the result is
\begin{equation}
    Z_{LS} \: = \: -\frac{M_p f_{LS}}{\sin\theta \sqrt{2}}
    \left(e^{-i\eta\ln\frac{1}{2}(1-\cos\theta)}-
        \frac{1}{2}(1-\cos\theta)\right)    \:\: ,
\end{equation}
where we have defined
\begin{equation}
   f_{LS} \: = \: -\frac{\alpha}{4M_{p}^{2}}(8\mu_{p}-2) \:\: .
\end{equation}
There is a similar contribution $-Z_{LS}$ to
$\langle10a|M^C_{C+N}(\theta)|11a\rangle$.
The corresponding result for $pp$ scattering~\cite{Knu78}
is the properly symmetrized version of this expression.
The partial-wave decomposition of $M^{C+M\!M}_{C+M\!M+N}(\theta)$,
the nuclear scattering amplitude in the presence of the
Coulomb force and the magnetic-moment interaction, is similar
to Eqn.~(\ref{eq:ampl}), but there now appears the following
$S$ matrix
\begin{equation}
   S_{C+M\!M+N} \: = \: S_C^{1/2}
   \left(S^C_{C+M\!M}\right)^{1/2} \, S^{C+M\!M}_{C+M\!M+N} \,
   \left(S^C_{C+M\!M}\right)^{1/2} S_C^{1/2} \:\: .
\end{equation}
Since the magnetic-moment interaction contains
a tensor part, the matrix $S^C_{C+M\!M}$ is not diagonal
in orbital angular momentum. However, the square root of this
matrix is still well-defined. What it all comes down
to, is to rewrite the $S$ matrix in the following manner
in order to split off the Coulomb amplitude $M_{C}$ and the amplitude
$M^{C}_{C+M\!M}$ due to the magnetic-moment interaction in the
presence of the Coulomb force
\begin{eqnarray}
   S_{C+M\!M+N}-1 & = & \left(S_C-1\right) +
   S_C^{1/2}\left(S^C_{C+M\!M}-1\right)S_C^{1/2} +
   \nonumber \\ &  & S_C^{1/2}\left(S^C_{C+M\!M}\right)^{1/2}
   \left(S^{C+M\!M}_{C+M\!M+N}-1\right)
   \left(S^C_{C+M\!M}\right)^{1/2}S_C^{1/2} \:\: ,
\end{eqnarray}
Other electromagnetic effects, like vacuum polarization,
can be treated in a similar way. For a more extensive discussion
we refer to Ref.~\cite{Sto90}. Because of the long range of the
magnetic-moment interaction the matrix elements of
$S_{C+M\!M+N}^{C+M\!M}$ are hard to calculate. We will use the
approximation
\begin{equation}
   S_{C+M\!M+N}^{C+M\!M} \approx S_{C+N}^{C}\ .
\end{equation}
Exactly the same approximation is made in the Nijmegen $N\!N$
partial-wave analyses~\cite{Sto90}.

\section{THE PHASE SHIFTS}
In this section we give the parametrization of the $\overline{N}\!N$
$S$ matrix in terms of phase-shift and inelasticity parameters.
We first seek guidance in the way this is done in analyses
of $N\!N$ scattering below the pion-production
threshold where the $S$ matrix is unitary and symmetric.
The symmetry of the $S$ matrix is a consequence
of time-reversal invariance which allows one to choose
the phases of the in- and out-states such that the
coupled-channels potential matrix, and thus the $S$ matrix,
is symmetric.
If there is conservation of flux, the $S$ matrix is unitary.

The phase shift for uncoupled partial waves with $\ell=J$, $s=0,1$
is defined by parametrizing the $1 \times 1$ $S$ matrix as
\begin{equation}
   S^J \: = \: \exp(2i\delta) \:\: .
\end{equation}
One usually denotes the different phase shifts by
$\delta_{\ell}$ for singlet $s=0$ waves, and by
$\delta_{\ell,J}$ for triplet $s=1$ waves.
For the partial waves with $\ell=J \pm 1 $, $s=1$
coupled by a tensor force one writes the $2 \times 2$
$S$ matrix in terms of two phase shifts
$\delta_{J-1,J}$, $\delta_{J+1,J}$ and one mixing
parameter $\varepsilon_J$. A popular
parametrization is the ``eigenphase'' convention
of Blatt and Biedenharn~\cite{Bla52}, in which
the symmetric $S$ matrix is diagonalized by way of a rotation
\begin{equation}
   S^J \: = \: \exp(-i\varepsilon_J\sigma_y) \,
               \exp(2i\delta) \,
               \exp(i\varepsilon_J\sigma_y) \:\: ,
\end{equation}
where
\begin{equation}
        \delta \: = \:
   \left( \begin{array}{cc}
     \delta_{J-1,J} &       0        \\
           0        & \delta_{J+1,J}
   \end{array} \right) \:\: .
\end{equation}
More often one uses the ``bar-phase''
convention of Stapp, Ypsilantis, and Metropolis~\cite{Sta57},
in which
\begin{equation}
   S^J \: = \: \exp(i\overline{\delta}) \,
               \exp(2i\overline{\varepsilon}_{J}\sigma_x) \,
               \exp(i\overline{\delta}) \:\: .
\end{equation}
An advantage of the ``bar-phase'' convention is that the
parameters go to zero when the interaction vanishes, unlike
the mixing parameter in the ``eigenphase'' convention.
Only in the former case is the mixing parameter a measure
of the strength of the off-diagonal tensor force.
We will use the ``bar-phase'' parametrization for the three
elastic parameters, since this is at present the common
choice in analyses of $N\!N$ scattering.

In the presence of coupling to annihilation channels
the $S$ matrix describing $\overline{N}\!N$ scattering
is only a submatrix of the much larger coupled-channels
$S$ matrix, and is therefore still symmetric, but no longer unitary.
This doubles the number of parameters needed.
The parametrization of the $S$ matrix
for uncoupled partial waves with $\ell=J$, $s=0,1$
in the presence of annihilation requires two parameters.
One writes
\begin{equation}
   S^J \: = \: \eta \, \exp(2i\delta) \:\: ,
\end{equation}
with $|\eta| \leq 1$.
To denote the inelasticities for the different partial waves we will
use a similar notation as for the phase shifts:
$\eta_{\ell}$ for singlet $s=0$ waves, and
$\eta_{\ell,J}$ for triplet $s=1$ waves.
For the coupled partial waves with $\ell=J \pm 1$, $s=1$
six parameters are needed, and it is not so easy to think of
a convenient parametrization which satisfies all constraints
from unitarity, is completely general, and free from
non-trivial ambiguities. Fortunately, the essential
work has already been done by Bryan~\cite{Bry81}
and others~\cite{Spr82,Kla83,Spr85}
in the case of $N\!N$ scattering above the
pion-production channel. Bryan generalizes the ``bar-phase''
convention by writing
\begin{equation}
   S^J \: = \: \exp(i\overline{\delta}) \,
               \exp(i\overline{\varepsilon}_{J}\sigma_x) \,\,
       H^J  \, \exp(i\overline{\varepsilon}_{J}\sigma_x) \,
               \exp(i\overline{\delta}) \:\: ,  \label{eq:Bryan}
\end{equation}
where $H$ is a three-parameter real and symmetric matrix
representing inelasticity. Bryan calls this matrix $N$, but
we use $H$ (capital $\eta$) to stress the analogy with
the case of uncoupled partial waves written in the form
\begin{equation}
   S^J \: = \: \exp(i\overline{\delta})
               \, \eta \, \exp(i\overline{\delta}) \:\: .
\end{equation}
If the inelasticity vanishes, then $H$ tends to the unit matrix,
and the ``bar-phase'' parametrization is recovered.
There are several nice ways to parametrize $H$, but we find
it convenient to follow Klarsfeld~\cite{Kla83},
who diagonalizes $H$ in Blatt-Biedenharn fashion
\begin{equation}
   H^J \: = \: \exp(-i\omega_J\sigma_y) \,
               \left( \begin{array}{cc}
                      \eta_{J-1,J} &       0        \\
                          0        &   \eta_{J+1,J}
                      \end{array} \right) \,
               \exp(i\omega_J\sigma_y) \:\: ,
\end{equation}
where the diagonal matrix contains the ``eigeninelasticities''
$\eta_{J-1,J}$ and $\eta_{J+1,J}$.
In this way the partial-wave annihilation cross section depends
only on the ``eigeninelasticities,'' since it is proportional to
\begin{equation}
   {\rm Tr} \, (1-H^2) \: = \: 2-\eta_{J-1,J}^2-\eta_{J+1,J}^2 \:\: .
\end{equation}
If the phase parameters are actually searched for on a computer,
it is better to write all inelasticities as
$\eta=\cos2\rho$. In this way, all parameters are real and unbounded.
The mixing parameter $\omega_J$ is finite
as the inelasticity vanishes, just as the mixing parameter
$\varepsilon_J$ in the ``eigenphase'' convention tends to a finite
value when the off-diagonal tensor force goes to zero.
Sprung~\cite{Spr85} has extended the Bryan parametrization
to allow the use of ``eigenphases.'' In this case one writes
\begin{equation}
   S^J \: = \: \exp(-i\varepsilon_J\sigma_y) \,
               \exp(i\delta) \,\,
        H^J \, \exp(i\delta) \,
               \exp(i\varepsilon_J\sigma_y) \:\: .
\end{equation}
The matrix $H$ in that case is, in general, not equal to the matrix
$H$ in Eqn.~(\ref{eq:Bryan}).
As stated above, we will use the ``bar-phases.''
The algorithm, due to Bryan~\cite{Bry81}, to extract the phase
parameters and inelasticities from the $S$ matrix
presented in numerical form is reviewed in the Appendix.

\section{STATISTICS}
Statistics is an essential ingredient in analyses of large
amounts of scattering data. The theoretical predictions
are compared with the experimental data using a least-squares
fitting procedure in which the model parameters are adjusted
to the data. During this process one continually
scrutinizes the data and passes sentence on the
quality of different sets, which sometimes have to be
rejected on the basis of statistical criteria. This goes on
until a final verdict is reached and the building of the
data set is completed. In this section this procedure is
outlined and the statistical tools required for our purpose
are presented. A more exhaustive treatment can be found
in Ref.~\cite{Ber88}.

We start by assuming for the moment that the measurements have no
normalization uncertainties and that no other type
of systematic error is present. In a certain
experiment, denoted by a subscript $a$, one has measured $N_{a}$ data
points. Each such measurement with statistical error is denoted by
$E_{a,i}\pm\varepsilon_{a,i}$ ($i=1,\ldots,N_{a}$). The model
prediction for a certain data point is given as $M_{a,i}({\bf p})$,
where the  model parameters
are arranged in a vector $\bf p$, with entries $p_{\alpha}$
($\alpha=1,\ldots,N_{\rm par}$).
The parameters are adjusted to the data by
minimizing the $\chi^2$-function
\begin{equation}
   \chi^2({\bf p}) \: = \: \sum_{a} \chi^2_{a}({\bf p}) \: = \:
   \sum_{a}\sum_{i=1}^{N_a} \left[ \frac{M_{a,i}({\bf p})-E_{a,i}}
   {\varepsilon_{a,i}} \right]^2 \:\:
\end{equation}
with respect to all parameters.

In practice, however, measurements usually do have an overall normalization
uncertainty, specified by the experimentalists. These errors
can be taken care of by introducing for each group of data
a normalization parameter $\nu_a$ with error $\varepsilon_{a,0}$.
The normalization of a certain group is then given as
$\nu_a=1\pm\varepsilon_{a,0}$. This means essentially that
for each group with a finite normalization uncertainty
another free parameter $\nu_a$ has been introduced, to be
determined in the fit, as well as an additional datum 1 with
error $\varepsilon_{a,0}$. Since we want to restrict the
parameter space to the model parameters only, we employ
the following trick to take the normalization parameters
into account. We redefine the $\chi^2$-function as follows
\begin{equation}
   \chi^2({\bf p}) \: = \: \sum_{a} \chi^2_{a}({\bf p}) \: = \:
   \sum_{a}{\rm min}\sum_{i=1}^{N_a}
   \left[ \frac{\nu_a M_{a,i}({\bf p})-E_{a,i}}
   {\varepsilon_{a,i}} \right]^2 + \left[ \frac{\nu_a-1}{\varepsilon_{a,0}}
   \right]^2 \:\: .    \label{eq:chi}
\end{equation}
In this way, the normalizations are adjusted trivially, by
minimizing in each iteration a quadratic function. In case the
normalization of an experiment is completely unknown,
$\varepsilon_{a,0}=\infty$ and we
can remove the second term on the right-hand
side of Eqn.~(\ref{eq:chi}). The corresponding normalization $\nu_a$
is determined in the fit. We say the normalization is ``floated.''
If the normalization is exactly known we fix the
normalization at $\nu_a=1$ and we can remove again the second term.
Angle-dependent normalization errors can be treated in a similar
manner. The minimum value $\chi^2_{\rm min}=
\chi^2({\bf p})\left|_{{\bf p}={\bf p}_{\rm min}}\right.$
is reached when
\begin{equation}
 \partial\chi^2({\bf p}) / \partial p_{\alpha}
   \: \equiv \: 0 \:\: ,
\end{equation}
for all values of $\alpha$. At this point one
defines the error matrix $E$ of the parameters as
\begin{equation}
   (E^{-1})_{\alpha\beta} \: = \: \frac{1}{2} \left.
 \partial^2\chi^2({\bf p}) / \partial p_{\alpha}\partial p_{\beta}
   \right|_{{\bf p}={\bf p}_{\rm min}} \:\: .
\end{equation}
The standard error on the
parameter $p_{\alpha}$ is $(E_{\alpha\alpha})^{1/2}$. Assuming
that the $\chi^2$-function is quadratic near its minimum, one can
show that this is the variation in $p_{\alpha}$ that gives a rise
$\Delta\chi^2=1$ in $\chi^2_{\rm min}$, when
the remaining parameters are refitted~\cite{Arn66}.

According to this discussion the following integers can be
defined. In the fit one must determine $N_{\rm n}$ normalization
parameters as well as $N_{\rm par}$ model parameters.
The actual number of free parameters is therefore
$N_{\rm fp}=N_{\rm par}+N_{\rm n}$. Of these $N_{\rm n}$
normalization parameters $N_{\rm ne}$ have a finite
normalization error and the rest $N_{\rm nf}=N_{\rm n}-N_{\rm ne}$
is the number of ``floated'' normalizations. The total
number of experimental scattering observables is $N_{\rm obs}$
and the actual number of scattering data is
$N_{\rm data}=N_{\rm obs}+N_{\rm ne}$. The number of degrees
of freedom is $N_{\rm df}=N_{\rm data}-N_{\rm fp}=
N_{\rm obs}-N_{\rm par}-N_{\rm nf}$.

In the process of screening the database, we employ
certain rejection criteria to remove data points that spoil
the statistical quality of the data set, for instance
by underestimated statistical errors or by unspecified
systematic errors. When statistical errors are underestimated,
data pretend to give more information than they actually do,
causing false results. By systematic errors we mean those errors
that do not average to zero when the measurement is repeated
many times, causing them to have a correlated effect
on the data. When not treated correctly, systematic errors
can bias the values of the model parameters and therefore also of the phase
parameter. A detailed discussion of this point
can again be found in Ref.~\cite{Ber88}. Of course,
we only reject data if there is conclusive evidence
against them. The rejection criteria are constructed in such a
way that in a purely statistical ensemble they have a very
small chance to occur. The following rejection criteria
have been employed by us.

$(i)$ An individual data point $E_{a,i}$ that has
$\chi^2_{a,i}>9$ is rejected.
%Such outliers can be seen
%as resulting from a faint broad background added to the
%probability distribution of the data.
Rejection of such outliers
will give more accurate values for the model parameters.
This rejection criterium corresponds to the usual
three-standard-deviation rule. It implies that a correct
datum has at most a chance of 0.27\% to be rejected.
The same criterium applies to an experimental normalization.
If it contributes more than 9 to $\chi^2_{\rm min}$,
this datum is rejected and the normalization is floated.

$(ii)$ A group of $N_a$ data is rejected if its $\chi^2_a$
in the multienergy fit is less than a minimum
$\chi^2_{\rm low}(N_a)$. The values for $\chi^2_{\rm low}$
can be found in Ref.~\cite{Ber88}. They are constructed
such that again the chance for a correct group to be rejected
is at most 0.27\%. This rejection criterium is used
as a means to avoid systematic errors.
Very low values of $\chi^2$
are very probably caused by systematic errors present in the
data and are \underline{not} a virtue of a model. This criterium
generally does not seem to be appreciated by the uninitiated in
statistical methods.

$(iii)$ A group of $N_a$ data is also rejected if its
$\chi^2_a$ in the multienergy fit exceeds a maximum
$\beta\chi^2_{\rm high}(N_a)$, where
$\beta=N_{\rm df}/(N_{\rm df}+N_{\rm par})$.
The values for $\chi^2_{\rm high}$ can also be found
in Ref.~\cite{Ber88}.

\section{THE NIJMEGEN 1993 ANTIPROTON-PROTON DATABASE}
\subsection{Set-up of the 1993 database}
In order to perform a $\overline{p}p$ PWA, we first
had to put together a statistically sound data set.
In $N\!N$ analyses, one has over 30 years
experience with the data, and by now a proper database
is more-or-less agreed upon. In the $\overline{p}p$ case,
we must start from scratch, but fortunately we can use the
experience of the $N\!N$ analyses to arrive at an acceptable
$\overline{p}p$ database.
Let us summarize the general features of our database.
We will include in our $\overline{p}p$ database all available
$\overline{p}p$ scattering data below antiproton laboratory momentum
$p_{\mbox{\scriptsize lab}}=925$ MeV/c published in a regular
physics journal since 1968. We do not take into
account data published only in conference abstracts,
in conference proceedings, and/or theses.
The reason we restrict ourselves to this momentum range is
that it corresponds roughly to the energy range
of the $N\!N$ partial-wave analyses
below kinetic energy $T_{\mbox{\scriptsize lab}}=350$ MeV
(momentum 925 MeV/c corresponds to $T_{\mbox{\scriptsize lab}}=379$ MeV)
and that we want to include
the accurate backward elastic cross sections between 406 and
922 MeV/c taken by Alston-Garnjost {\it et al}.~\cite{Als79}.
Low-energy data, below $p_{\rm lab} = 175$ MeV/c, so
$T_{\mbox{\scriptsize lab}} \approx 15$ MeV, are of course almost
nonexistent here. Only scattering observables are analyzed,
other ``data'' like for instance the real-to-imaginary ratio
of the forward scattering amplitude or the slope of the
$\overline{p}p$ forward nuclear amplitude are omitted,
since the extraction of these quantities from the data is
model dependent. A summary of the Nijmegen 1993 $\overline{p}p$
database can be found in Table~\ref{tab:data}.

Several experiments~\cite{Car74,Cha76,Sak79} have reported
a resonant structure in the antiproton-proton total cross section
near 490 MeV/c without agreeing, however, on the exact position,
strength, and width of a possible resonance. We do not include
these data in the PWA, since more recent and accurate
measurements of total cross
sections~\cite{Kam80,Ham80b,Sum82,Nak84a,Clo84}
have convincingly ruled out the existence of this type of
rather broad resonance.
(A more narrow type of resonance, however, is not ruled out
by the existing data. In fact, there is some statistical evidence
for a narrow structure in backward elastic cross sections
around 509 MeV/c~\cite{Tim85}. Probably, only an accurate
measurement of the backward charge-exchange cross section can
definitely settle this issue.)

High-quality total cross sections below 400 MeV/c have been
measured at LEAR by the PS172 collaboration~\cite{Bug87}.
The effects of the pure Coulomb force and Coulomb-nuclear
interference are not significant, except perhaps at low momenta
$\alt$ 300 MeV/c. This may be the reason why we had
to reject the data at the two lowest momenta.

In a number of
experiments~\cite{Cha76,Bru77,Ham80b,Jas81,Low81,Sai83,Bra85,Fra87}
the $\overline{p}p$ annihilation cross section into charged
mesons has been measured. We cannot include these data, not even
with a floated normalization, since the
momentum dependence of the cross section for annihilation
into neutral mesons is not known. At LEAR the total annihilation
cross section, including annihilation into neutral channels,
has been measured from 180 to 600 MeV/c by the PS173
group~\cite{Bru87,Bru90}. These last data are included.

We do not include integrated elastic-cross-section
data~\cite{Con68,Spe70,Cou77,Sak82,Kag87}, due to difficulties
in a proper treatment of Coulomb and Coulomb-nuclear interference.
Integrated charge-exchange cross sections, on the other
hand, are included in the database. %since there is no significant
%effect due to the initial-state Coulomb interaction.
In two experiments~\cite{Als75,Ham80a} this observable
was measured over a large momentum region. For the experiment
of Ref.~\cite{Ham80a} data points at the lowest momenta
are rejected since they are in conflict with more recent
measurements done at LEAR by the PS173 group~\cite{Bru86b}.
Our PWA clearly favors this last experiment.

In the pre-LEAR era a very large part of the experimental
data on $\overline{p}p$ scattering consisted of elastic
differential cross sections~\cite{Con68,Spe70,Koh72,Alb72,Eis76,Sak82}.
The most accurate data were those taken by Eisenhandler {\it et al}.
at 690, 790, and 860 MeV/c (and higher momenta)~\cite{Eis76}.
With the exception of the data at 194.8 MeV/c
of Ref.~\cite{Spe70} and the data at 910 MeV/c of Ref.~\cite{Alb72}
we find these pre-LEAR data to be consistent. An accurate measurement
of the backward elastic differential cross section at $\cos\theta$ =
$-$0.994 was done by Alston-Garnjost {\it et al}.~\cite{Als79} between 406
and 922 MeV/c. Since 1983, differential cross sections on $\overline{p}p
\rightarrow \overline{p}p$ have been measured
by different groups at LEAR and at KEK. For an extensive
discussion of these data we refer to the next subsection.

A number of pre-LEAR experiments determined the
elastic differential cross section at forward scattering angles
in order to study the Coulomb-nuclear--interference
region~\cite{Kas78,Iwa81,Cre83,Ash85}. At LEAR this was done by the
collaborations PS172~\cite{Lin87,Sch89} and PS173~\cite{Bru85}.
In general, all these data are well described in the PWA,
although in some cases data points in the
extreme-forward region had to be rejected since they
are contaminated by multiple Coulomb scattering in the target
(so-called Moli\`ere scattering).

Some of the most important results coming from LEAR
are the high-quality elastic analyzing-power $A_y$ (polarization)
data, measured by the PS172~\cite{Kun88,Kun89} and
PS198~\cite{Ber89,Per91} collaborations at a number of
energies above 439 MeV/c. Since in the pre-LEAR era
only very few and inaccurate data points
existed~\cite{Alb72,Ohs73,Kim82}, these experiments
mean real progress. We find the data of PS172 and PS198
to be consistent, except for their normalizations.
In our PWA we float the normalizations of the
$A_y$ data at 497, 523, and 679 MeV/c. PS172 has also obtained
the first (not very accurate) results on the elastic $D_{yy}$
depolarization~\cite{Kun91}. For these data we do not take a
normalization error into account, in view of the large
error bars.

Before the advent of LEAR in 1983, also charge-exchange
differential observables were scarce. Some differential cross
sections existed~\cite{Biz68,Koh72,Bog76,Tsu83,Ban85}, but these
were not very accurate. Since 1984, however, the situation has
improved enormously. High-quality differential cross
sections have been obtained at KEK between 392 and 781
MeV/c~\cite{Nak84b}, and at LEAR by the PS173 group
at low momenta between 183 and 590 MeV/c~\cite{Bru86b}.
One of the most important experiments at LEAR has been
PS199 whose goal was to study the spin structure of the
charge-exchange reaction. So far, it has obtained
very accurate data on the differential charge-exchange
cross section at 693 MeV/c and very important data
on the charge-exchange analyzing power $A_y$ between 546 and 875
MeV/c~\cite{Bir90,Bir91}. More results from PS199
can be expected in the near future. Very recently,
an new LEAR experiment called PS206~\cite{Mac92} has been approved
that will further study the charge-exchange reaction.

\subsection{Flaws in elastic differential cross-section data}
Since 1983 several experiments have
measured the elastic differential cross section at different
momenta. PS173 at LEAR measured this observable at low momenta
between 181 and 590 MeV/c~\cite{Bru86a,Bru91a}. It was subsequently
measured at KEK between 392 and 781 MeV/c~\cite{Kag87}, by PS172
at 679, 783, and 886 MeV/c~\cite{Kun89} (and at higher energies),
and by PS198 at 439, 544, and 697 MeV/c~\cite{Ber89,Per91}.
Unfortunately, these different experiments do not appear to be consistent
with each other, nor with the accurate pre-LEAR data of Eisenhandler
{\it et al}.~\cite{Eis76} which are described very well in
the PWA. In fact, this is the most serious
flaw in the $\overline{p}p$ database and a major
obstacle in fitting potential models~\cite{Pig91}.
It is highly probable that some of these data contain unspecified
systematic errors, or that the statistical errors have been
underestimated.

Since we are talking about a total of 540 cross
sections (not counting the Eisenhandler data)
we decided to put some effort in trying to determine what is wrong
with these data, instead of rejecting them outright. This turned out
to be very difficult. The discussion about statistical and systematic
errors in most of the original papers can be called marginal at best.
Sometimes results from Legendre fits are presented, but in several cases
it turns out that the $\chi^2_{\rm min}$ of these fits is much too high
for a statistical data set. The procedure that we followed to
examine these data was as follows.
We fitted each differential cross section with
\[   {\rm d}\sigma/{\rm d}\Omega \: = \:
  \sum_{\ell=0}^{\ell_{\rm max}} \: a_{\ell} P_{\ell}(\cos\theta) \:\: . \]
$\ell_{\rm max}$ was determined by the requirement that the error on
the corresponding coefficient $a_{\ell}$ was smaller than the coefficient
itself. In the ideal situation this would give a fit with
$\chi^2_{\rm min}/N_{\rm df} \approx 1.0$. However, for few
groups this was actually the case. We then enlarged the errors
by adding a point-to-point systematic error in quadrature to the
quoted statistical errors. The amount of error added was determined by
the requirement that now indeed $\chi^2_{\rm min} \approx N_{\rm df}$.
Once for a group this result was approached, outliers were removed
(three-standard-deviation rule). For some groups special
measures had to be taken, which we will discuss now. The results of
these investigations are summarized in Table~\ref{tab:flaws}.

The data measured by PS173~\cite{Bru86a,Bru91a} are at the most forward
angles contaminated by effects due to multiple scattering in the target
(Moli\`ere scattering). Since the experimentalists do not present their
data corrected for these effects, and since these corrections depend
on details of the target used, there is no way for us to take these effects
into account. Consequently, the only sensible thing to do is to reject
the data at angles where multiple-scattering effects are believed to
be seen. These data were used to extract the real-to-imaginary ratio
of the forward scattering amplitude~\cite{Bru85}. It is perhaps remarkable
that in Ref.~\cite{Bru85} values for this ratio were presented at more
momenta than the four for which the corresponding differential cross
sections were presented in Refs.~\cite{Bru86a,Bru91a}. These remaining
cross sections have never been published. After removing the points
at forword angles, difficulties remain. At 287 MeV/c it is necessary
to reject five individual data points and add a 5\% point-to-point
error. At 505 MeV/c two outliers must be removed and at 590 MeV/c
a 3\% error must be added.

The data taken at KEK~\cite{Kag87}
present serious difficulties of a different kind.
The only manner to achieve satisfactory results in a Legendre fit
seemed to be to either reject a large number of data points or to
enlarge all errors by a significant amount ($\approx$ $6-7$\%).
However, improvement could be obtained in the following manner.
It turned out that most difficulties resided in the data taken in
the ``one-prong'' region~\cite{Kag87}, i.e. at forward and backward
angles. The data at intermediate angles, the ``two-prong'' region,
appear to be less troublesome. In fact, in Ref.~\cite{Kag87} different
systematic errors are quoted for these two regions, although the
experimentalists perform Legendre fits to the data as one group.
When we split the groups at 490, 591, 689, and 781 MeV/c into two
different parts, it suffices to enlarge the statistical errors by a
smaller amount, and in some ``two-prong'' regions the data need no
corrections at all.

The description of the data taken by PS172~\cite{Kun89} improves much
when the points at the most forward and at the most backward angle
at removed at all momenta (F. Bradamante, private communication).
At 886 MeV/c three additional outliers have to be rejected.
Especially for these groups there remains a big problem with the
normalization~\cite{Kun89} (see below). For the data
at 697 MeV/c~\cite{Ber89} from the PS198 group the statistical
errors may have been underestimated. We had to enlarge
these by adding a 6\% point-to-point error.

After all these corrections have been applied to the different groups
the next step is to see to what extent the Legendre coefficients
at comparable momenta are consistent. These coefficients are presented
in Table~\ref{tab:Legendre}. It is immediately clear that significant
problems occur in the normalization $a_0$ of the different groups
(which is why we present the ``renormalized'' Legendre coefficients
$a_{\ell}/a_0$ for $\ell>0$). This is especially true for the PS172 data.
The probable reason~\cite{Kun89} for the difficulties is that data could
only be taken in a very limited angular region, so that only a small
fraction $\approx$ 5\% of the cross section is detected. Properly
normalizing the data is then especially difficult. These data would
have to be treated with a ``floated'' normalization. Certainly a 10\%
normalization uncertainty as suggested in Ref.~\cite{Kun89} is not
sufficient. Since also the KEK data in the ``two-prong'' region
cover a limited angular region it would be a good idea to float
these normalizations as well, since they are not consistent with
the normalizations of the data in the ``one-prong'' regions.
Apart from these normalization problems it is clear from a study
of Table~\ref{tab:Legendre} that it is very unlikely that the
different experiments are consistent. Certainly, the PS173 data at 590
MeV/c are not compatible with the KEK data at 591 MeV/c; the PS172
data at 783 MeV/c are not compatible with the KEK data at 781 MeV/c,
nor with the Eisenhandler data at 790 MeV/c. The PS198 data at 439 MeV/c
do not appear to intrapolate between the KEK data at 392 and 490 MeV/c,
etcetera. So, although many beautiful data have come out of LEAR and KEK,
unfortunately elastic differential cross sections are not among them.

To summarize, at present we are prejudiced in favor of the pre-LEAR
data by Eisenhandler {\it et al}.~\cite{Eis76}, although we cannot
exclude completely the possibility that these data contain unspecified
systematic errors and that one of the LEAR or (more unlikely) the KEK
experiments is correct after all. Preliminary study showed that
we could obtain a reasonable fit to the LEAR data from PS172 and
PS198, but only at the cost of rejecting the pre-LEAR data of
Eisenhandler {\it et al}.~\cite{Eis76} and of Sakamoto
{\it et al}.~\cite{Sak82}. And still, the problems with the
data from PS173 and KEK would remain. Further investigation is
required before such a drastic step will be taken.
Obviously, a dedicated new experiment that might shed some
light on this issue would be highly welcome.

\section{RESULTS}
After deciding on the final content of the Nijmegen 1993
$\overline{p}p$ database to be used in the PWA, the free
$P$-matrix parameters are fitted to these data\footnote{For
the sake of completeness we mention that at the time of our
final fit, the data of Refs.~\cite{Bog76,Ban85,Ash85} were not
available to us. Also, we were at that time not aware of
the existence of the data of Refs.~\cite{Kam80,Nak84a}.}.
In Table~\ref{tab:data} we present an overview of
the results of this fit for all the
$\overline{p}p$ scattering data. In this Table
one can find the values for $\chi^2_{\rm min}$ for each
individual group, the normalization predicted by the
PWA, and the data points and groups that have been
rejected by us on statistical grounds. The total number of
scattering observables that are rejected is 204,
not including the problematic LEAR and KEK elastic differential
cross sections discussed in the previous Section. Three
normalization data are rejected. We reject three
groups because of an improbable high $\chi^2_{\rm min}$,
and two groups because of an improbable low $\chi^2_{\rm min}$.
In the final fit we have (in the notation of Sect. VII)
\[
        N_{\rm obs} = 3543  \: , \:\:\:
        N_{\rm n}   = 113   \: , \:\:\:
        N_{\rm ne}  = 103   \: , \:\:\:
        N_{\rm par} = 30    \: ,
\]
so that
\[
        N_{\rm data} = 3646  \: , \:\:\:
        N_{\rm fp}  = 143   \: , \:\:\:
        N_{\rm nf}  = 10    \: , \:\:\:
        N_{\rm df}  = 3503  \: .
\]
When the data set is a perfect statistical ensemble and
when the model is totally correct one expects
\[  \langle \chi^2_{\rm min}/N_{\rm df} \rangle
    \: = \: 1.000 \pm 0.024 \:\: . \]
Our best fit to the final 1993 database results in
\[     \chi^2_{\rm min} \: = \: 3801.0 \:\: ,   \]
corresponding to
\[     \chi^2_{\rm min}/N_{\rm data} \: = \: 1.043  \:\:\:\:\: {\rm and}
   \:\:\:\:\: \chi^2_{\rm min}/N_{\rm df}  \: = \: 1.085 \:\: .   \]
The 3543 scattering observables contribute 3700.9 to $\chi^2_{\rm min}$,
which means that the 103 normalization data contribute the remaining
100.1 of $\chi^2_{\rm min}$.

In order to get a feeling for some of these numbers, let
us compare them to the $N\!N$ case. In the Nijmegen $N\!N$ PWA
the database contains $N_{\rm data}=4301$ $N\!N$ scattering data,
which are described with $\chi^2_{\rm min}=4263.8$ or
$\chi^2_{\rm min}/N_{\rm data}=0.991$~\cite{Klo93}. The
number of model parameters needed in our case is 30, which is a
reasonable number, in view of the fact that 21 parameters were used
in the Nijmegen $pp$ PWA and an additional 18 in the $np$ PWA.

The values for the $P$-matrix parameters and their errors
are tabulated in Table~\ref{tab:pars}. The parameters for the
higher partial waves are the same as the corresponding state
given in this Table. For instance, the $^1F_3$ wave has the
same parameters as the $^1D_2$ wave, the $^3F_3$ wave the
same as the $^3D_2$ wave, the $^3G_3$ the same as the
$^3F_2$, and so on. This is just a convenient prescription.
For higher partial waves the short-range parametrization
is irrelevant due to the centrifugal barrier and the dynamics is
completely determined by the potential tail. As explained in
Section III, these parameters correspond to a spin-dependent optical
potential for the short-range interaction. It can be seen
that even if one introduces for each partial wave of specific
isospin a simple complex spherical-well potential, then
still by no means all of these parameters can be determined
from the existing data. This is in striking
contrast with for instance the $pp$ analysis where one needs
4 parameters already to describe the $^1S_0$ channel.
The reason is that the $pp$ $^1S_0$ phase shift is very accurately
known at very low energies (below 3 MeV). A proper description
of the $P$ waves in the $pp$ case also requires more than one
parameter for each wave. In our $\overline{p}p$ case,
no single partial wave (of specific isospin) needs more than
one parameter for the real part of the short-range interaction,
but on the other hand much more partial waves contribute
significantly to the scattering process. This is especially
true for the charge-exchange reaction, as can be seen
from Table~\ref{tab:partxs} where we give the partial-wave
cross sections for the elastic and charge-exchange reaction
at a number of momenta, as well as the total and annihilation
cross section. Compared to nucleon-nucleon scattering,
one notes a large contribution of $P$ and $D$ waves to the
cross sections already at low momenta. The reason for this
is the greater strength of the antinucleon-nucleon potentials,
especially the central and tensor potentials.

In Figure~\ref{fig:stot} we show the results for the
total and annihilation cross sections as a function of
momentum. The high-quality data are from the LEAR experiments
PS172~\cite{Bug87,Clo84} (total cross sections) and
PS173~\cite{Bru87,Bru90} (annihilation cross sections).
We calculate total cross sections with the optical
theorem. Obviously, these cross sections can only be
compared to the experimental data, when the effects
of the Coulomb interaction can be neglected. Except
for the lowest momenta ($p_{\rm lab}<$ 200 MeV/c)
this is probably a good approximation.
An example of the fit to the differential cross sections
from Eisenhandler {\it et al}.~\cite{Eis76} is shown in
Figure~\ref{fig:xsel}. In Figure~\ref{fig:xsbw} the fit
to the backward elastic cross sections ($\cos\theta=-0.994$)
from Ref.~\cite{Als79} is demonstrated. Assuming that
these data are correct, it can be seen from this Figure
that there appears to be room for improvement at the lowest
momenta. This is precisely the momentum region
($p_{\rm lab}\approx$ 509 MeV/c) where some statistical
evidence for a resonance was found~\cite{Tim85}.
The differential cross section for charge-exchange
scattering is given in Figure~\ref{fig:xsce}, compared
to recent high-quality data taken by PS199~\cite{Bir90}
at 693 MeV/c. This is one of the most constaining
experiments in the database. In order to fit this group
properly, orbital angular momenta up to $\ell=10$ must
be taken into account.
The cross section exhibits the typical dip-bump
structure at forward angles, which can be understood
as an interference effect between one-pion exchange and
a background~\cite{Lea76}. Unfortunately, no data
of similar quality have been taken in this dip-bump region.

In Figure~\ref{fig:ayel} we give the results for the
analyzing power (polarization) for elastic scattering,
compared with the recent data from PS172~\cite{Kun88}
and from PS198~\cite{Ber89,Per91}. The fits to
the analyzing-power data for the charge-exchange reaction
from PS199~\cite{Bir90,Bir91} are shown in Figure~\ref{fig:ayce}.
Finally, in Figure~\ref{fig:dyy} the prediction for the
depolarization for elastic
scattering at 783 MeV/c is shown compared to the data from
PS172~\cite{Kun91}. Only a few depolarization data exist:
one point at 679 MeV/c, three points at 783 MeV/c, and one point
at 886 MeV/c. These data points are not included in the fit
in view of the large error bars. This also means that
no normalization error is taken into account.

In Sect.~VI a formalism was proposed to extract phase-shift
parameters and inelasticities from the $S$ matrix for
antinucleon-nucleon scattering. It does not make much
sense to present all these phase shifts, inelasticities
and mixing parameters without a proper assessment of the
uncertainties (statistical errors). This, however, requires
a lot of work. Preliminary study shows
that the phase-shift parameters for the $^1S_0$ and
$^1P_1$ partial waves are not pinned down accurately at all above
$p_{\rm lab}\approx$ 400 MeV/c. On the other hand, a large number
of parameters appear to be very well determined by the existing data,
such as the $^3P_{0,1,2}$ and $^3D_{2,3}$ phase shifts and inelasticities,
and the $\varepsilon_{1,2}$ mixing parameters. All results
are available in numerical form from the authors.
An extensive discussion of the remaining uncertainties
in the predictions of the PWA and the phase-shift parameters
will be presented elsewhere.

In our 1991 preliminary PWA~\cite{Tim91b} we were able to
determine the charged-pion--nucleon coupling constant
$f^2_{\mbox{\scriptsize c}}$ from the data on the charge-exchange
reaction $\overline{p}p \rightarrow \overline{n}n$, in which
only isovector mesons can be exchanged. In this PWA we
analyzed 884 scattering observables between 400 and 950 MeV/c,
using 23 free parameters. We found
$f^2_{\mbox{\scriptsize c}}$ = 0.0751(17), at the pion pole.
The error is purely statistical. This value was in nice
agreement with other determinations of $f^2_{\mbox{\scriptsize c}}$
from $np$~\cite{Klo91} and $\pi^{\pm}p$
scattering data~\cite{Arn90}. These results
provided strong evidence for an approximate
charge-independent pion-nucleon coupling constant, since
they were consistent with the value for $f^2_{\mbox{\scriptsize p}}$
found in the Nijmegen $pp$ PWA~\cite{Ber90}.

Since this preliminary analysis, more high-quality
analyzing-power data for the charge-exchange reaction
have become available from PS199~\cite{Bir91}.
We have repeated the determination of $f^2_{\mbox{\scriptsize c}}$,
but this time from the complete 1993 Nijmegen database.
The coupling constants of the neutral pion were kept at the
value of $f^2_{\mbox{\scriptsize p}}$ =
$f^2_{\mbox{\scriptsize n}}$ = 0.0745.
Since $f^2_{\mbox{\scriptsize p}}$ is determined by the data on
elastic scattering $\overline{p}p \rightarrow \overline{p}p$,
and $f^2_{\mbox{\scriptsize c}}$ by the data on charge-exchange
scattering $\overline{p}p \rightarrow \overline{n}n$, one
expects that the correlation between the two is small.
Adding $f^2_{\mbox{\scriptsize c}}$ as the 31st free
parameter, we now find
\[  f^2_{\mbox{\scriptsize c}} = 0.0732(11) \:\: , \]
at the pion pole.
This result supersedes our previous value from Ref.~\cite{Tim91b}.
Again, the error is of statistical origin only. In view of
the enormous amount of work involved, it is very
hard for us to make statements about possible systematic errors
on this result. In Ref.~\cite{Tim91b} we did demonstrate that
there were no systematic errors due to form-factor effects
or due to $\rho$(770) exchange. In the Nijmegen $pp$ PWA
systematic errors could be more thoroughly investigated and there
they were found to be small~\cite{Sto93}. Although in our
case the systematic errors are probably larger than for
the $pp$ case, it is very
encouraging that the available charge-exchange data pin
down the charged-pion coupling constant with a remarkably
small statistical error and that the result is consistent
with other determinations from $\pi^{\pm}p$~\cite{Arn90} and
$N\!N$ scattering~\cite{Klo91,Sto93}.
Very probably the new LEAR experiment PS206~\cite{Mac92} will
further constrain the charged-pion--nucleon coupling constant.

\section{SUMMARY OF CONCLUSIONS}
To summarize, we have performed an energy-dependent partial-wave
analysis of all antiproton-proton scattering data below
925 MeV/c antiproton momentum, published in a regular physics
journal since 1968. This is the first time such an analysis
has been attempted.
We have set up the Nijmegen 1993 $\overline{p}p$ database
by scrutinizing and passing sentence on all available
$\overline{p}p$ scattering observables below 925 MeV/c.
Serious problems were encountered with a set of 540 elastic
differential cross sections from LEAR and KEK. These data
were rejected, although further study is required here.
Of the remaining 3747 scattering observables 204
(5.4\%) were rejected on the grounds of sound statistical
criteria. We also rejected three normalization data.
The final database contains 3543 scattering
observables and 103 normalization data for a total
of 3646 scattering data. Using 30 free
parameters we obtain $\chi^2_{\rm min}$ = 3801.0
corresponding to $\chi^2_{\rm min}/N_{\rm data}$ = 1.043.
This shows that the tail of the charge-conjugated Nijmegen
potential is a realistic intermediate- and long-range
$\overline{p}p$ force. Data on the
charge-exchange reaction $\overline{p}p \rightarrow
\overline{n}n$ provide further evidence for a ``low''
and approximately charge-independent
pion-nucleon coupling constant $f^2_{N\!N\pi}$ $\approx$ 0.0745.
The present results will serve as a starting-point for future
investigations by our group of the antiproton-proton system,
and should be of help in planning further experiments at LEAR and
elsewhere.

\acknowledgments
We would like to thank the members of the PS172, PS173, PS198,
and PS199 LEAR collaborations at CERN for providing us with
their data and for their interest in our work. We also thank
Drs. B. Loiseau, S. Sakamoto, and M. Cresti for sending us data.
Conversations with Dr. V. Stoks and M. Rentmeester on details of
the Nijmegen nucleon-nucleon partial-wave analyses are
gratefully acknowledged. Part of this work was included
in the research program of the Stichting voor Fundamenteel
Onderzoek der Materie (FOM) with financial support from the
Nederlandse Organisatie voor Wetenschappelijk Onderzoek (NWO).

\appendix
\section*{FROM S MATRIX TO PHASE PARAMETERS}
In his second paper on the subject of parametrizing the $S$ matrix
for nucleon-nucleon scattering above the pion-production
threshold~\cite{Bry81}, Bryan has presented an easy
algorithm to extract the three parameters
$\overline{\delta}_{J-1,J}$, $\overline{\delta}_{J+1,J}$,
$\varepsilon_J$, and the three different elements of the
inelasticity matrix $H^J$ from the $2 \times 2$ $S$ matrix,
written as
\begin{equation}
   S^J \: = \: \exp(i\overline{\delta}) \,
               \exp(i\overline{\varepsilon}_{J}\sigma_x) \,\,
       H^J  \, \exp(i\overline{\varepsilon}_{J}\sigma_x) \,
               \exp(i\overline{\delta}) \:\: .
\end{equation}
For completeness we list the relevant expressions.
The derivation can be found in the paper by Bryan.
If the $S$ matrix is presented numerically as
\begin{equation}
        S^J   \: = \:
   \left( \begin{array}{cc}
     R_{11}\exp(2i\delta_{11}) & iR_{12}\exp(2i\delta_{12}) \\
    iR_{12}\exp(2i\delta_{12}) &  R_{22}\exp(2i\delta_{22})
   \end{array} \right) \:\: ,
\end{equation}
then the phase shifts
$\overline{\delta}_{J-1,J}$ and $\overline{\delta}_{J+1,J}$
can be obtained from the following two equations
\begin{eqnarray}
   \tan2(\theta_a+\theta_b) & = &
   \frac{R^2_{12}\sin2\delta}{R_{11}R_{22}+R^2_{12}\cos2\delta}
   \:\: , \\
   \tan(\theta_a-\theta_b) & = &
   \frac{R_{22}-R_{11}}{R_{11}+R_{22}}\tan(\theta_a+\theta_b)
   \:\: .
\end{eqnarray}
Here we defined the auxilary phases
\begin{eqnarray}
   \theta_{a} & \equiv & \delta_{11}-\overline{\delta}_{J-1,J}
   \:\: , \\
   \theta_{b} & \equiv & \delta_{22}-\overline{\delta}_{J+1,J}
   \:\: , \\
   \delta     & \equiv & \delta_{11}+\delta_{22}-2\delta_{12}
   \:\: .
\end{eqnarray}
The mixing parameter $\overline{\varepsilon}_J$ can
subsequently be calculated from
\begin{equation}
   \tan2\varepsilon \: = \:
   \frac{2R_{12}\cos(\theta_a+\theta_b-\delta)}
        {R_{11}\cos2\theta_a + R_{22}\cos2\theta_b} \:\: ,
\end{equation}
where $\varepsilon\equiv\overline{\varepsilon}_J$.
Next the elements of the matrix $H^J$ are isolated.
One finds
\begin{eqnarray}
   2\cos2\varepsilon \, H_{11} & = &
      R_{11}\cos2\theta_a (1+\cos2\varepsilon) +
      R_{22}\cos2\theta_b (1-\cos2\varepsilon) \:\: , \\
   2\cos2\varepsilon \, H_{22} & = &
      R_{11}\cos2\theta_a (1-\cos2\varepsilon) +
      R_{22}\cos2\theta_b (1+\cos2\varepsilon) \:\: , \\
    \cos2\varepsilon \, H_{12} & = &
      R_{12}\sin(\delta-\theta_a-\theta_b)  \:\: .
\end{eqnarray}
If the matrix $H^J$ is parametrized according to
Klarsfeld~\cite{Kla83}, the three inelastic parameters
$\eta_{J-1,J}$, $\eta_{J+1,J}$, and $\omega_J$
can be obtained from
\begin{eqnarray}
    \eta_{J-1,J} + \eta_{J+1,J} & = & {\rm Tr}  \, H^J \:\: , \\
       \eta_{J-1,J}\eta_{J+1,J} & = & {\rm det} \, H^J \:\: , \\
                  \tan2\omega_J & = & 2H_{12}/(H_{11}-H_{22})
    \:\: .
\end{eqnarray}

\newpage
\begin{figure}
\caption{Total and annihilation cross section as a function of
        momentum in antiproton-proton scattering. The $\sigma_{\rm tot}$
        data are from the LEAR experiment PS172~\protect\cite{Clo84,Bug87}
        and the $\sigma_{\rm ann}$ data are from experiment
        PS173~\protect\cite{Bru87,Bru90}. The curve from the PWA for
        $\sigma_{\rm tot}$ has $\chi^2_{\rm min}=88.4$ for 75 points,
        and the curve for $\sigma_{\rm ann}$ has $\chi^2_{\rm min}=65.3$
        for 52 points.}
\label{fig:stot}
\end{figure}

%\newpage
\begin{figure}
\caption{Differential cross section for elastic scattering at 790 MeV/c.
        The data are from Eisenhandler {\it et al}.~\protect\cite{Eis76}.
        The curve from the PWA has $\chi^2_{\rm min}=101.5$ for 95
        points.}
\label{fig:xsel}
\end{figure}

%\newpage
\begin{figure}
\caption{Elastic cross section at backward angle $\cos\theta=-0.994$
         as a function of momentum. The data are from Alston-Garnjost
         {\it et al}.~\protect\cite{Als79}. The curve from the PWA has
         $\chi^2_{\rm min}=36.7$ for 30 points.}
\label{fig:xsbw}
\end{figure}

%\newpage
\begin{figure}
\caption{Differential cross section for charge-exchange scattering
         at 693 MeV/c. The data are from the LEAR experiment
         PS199~\protect\cite{Bir90}. The curve from the PWA has
         $\chi^2_{\rm min}=39.3$ for 33 points.}
\label{fig:xsce}
\end{figure}

%\newpage
\begin{figure}
\caption{Analyzing power in elastic scattering at 544, 679, 783,
         and 886 MeV/c. The data are from the LEAR experiments
         PS172~\protect\cite{Kun88} and PS198~\protect\cite{Per91}.
         The curves from the PWA have $\chi^2_{\rm min}=$37.5, 23.1,
         30.1, and 38.5 for 27, 26, 29, and 34 points, respectively.}
\label{fig:ayel}
\end{figure}

%\newpage
\begin{figure}
\caption{Analyzing power in charge-exchange scattering at 546, 656,
         767, and 875 MeV/c. The data are from
         the LEAR experiment PS199~\protect\cite{Bir90,Bir91}.
         The curves from the PWA have $\chi^2_{\rm min}=$36.1, 21.9,
         25.5, and 20.0 for 23, 21, 22, and 23 points, respectively.}
\label{fig:ayce}
\end{figure}

%\newpage
\begin{figure}
\caption{Depolarization in elastic scattering at 783 MeV/c. The
         data are from the LEAR experiment PS172~\protect\cite{Kun91}.
         The curve from the PWA has $\chi^2_{\rm min}=4.9$ for 3
         points.}
\label{fig:dyy}
\end{figure}

\newpage
\widetext
\begin{table}
\caption{Reference table of antiproton-proton
         scattering data below 925 MeV/c.}
\begin{tabular}{cccccccc}
$p_{\rm lab}$& $\rm No.^a$ &
              &  Norm  &  Pred.    &          &           &          \\
(MeV/c) &  $\rm type^b$ & $\chi^2_{\rm min}$ &
 error & $\rm norm^c$ & $\rm Rejected^d$ & Ref. & Comm.   \\
\tableline
\dec 119.0-- &                           &
                           &     &       &      &             &       \\
\dec --923.0 &   50  $\sigma_{\rm ce}$   & \dec   29.2
                          &3--5\%& 1.098 & $\leq$385.0, \#=8
                                                &\cite{Ham80a}& k, m  \\
\dec 176.8-- &                           &
                           &     &       &      &             &       \\
\dec --396.1 &    5  $\sigma_{\rm ann}$  & \dec    7.5
                           &4.4\%& 0.943 & 176.8&\cite{Bru90} &       \\
\dec 181.0   &   46 d$\sigma_{\rm el}$   & \dec   48.7
                           &  5\%& 1.037 & $\geq$0.925, \#=6
                                         &\cite{Bru86a,Bru91a}& j, o  \\
\dec 183.0   &   13 d$\sigma_{\rm ce}$   & \dec    8.3
                           &  5\%& 0.976 & 0.940, $-$0.170,   &&      \\
                       &&&&& $-$0.574, $-$0.966 &\cite{Bru86b}&       \\
\dec 194.8   &   19 d$\sigma_{\rm el}$   &          .
                           &  4\%&  .    & all  &\cite{Spe70} &f, i, o \\
\dec 200.0-- &                           &
                           &     &       &      &             &       \\
\dec --588.2 &   48  $\sigma_{\rm ann}$  & \dec   57.8
                           &2.2\%& 0.978 &      &\cite{Bru90,Bru87}&  \\
\dec 221.9-- &                           &
                           &     &       &      &             &       \\
\dec --413.2 &   45  $\sigma_{\rm tot}$  & \dec   53.0
                           &0.9\%& 0.971 & 221.9, 229.6
                                                &\cite{Bug87} &       \\
\dec 233.0   &   54 d$\sigma_{\rm el}$   & \dec   87.6
                           &  5\%& 1.034 & $\geq$0.938, \#=6; &&      \\
                      &&&&& 0.764               &\cite{Lin87} &   j   \\
\dec 239.2   &   20 d$\sigma_{\rm el}$   & \dec   27.7
                           &  4\%& 1.089 &      &\cite{Spe70} &   o   \\
\dec 272.0   &   65 d$\sigma_{\rm el}$   & \dec   55.9
                           &  5\%& 1.055 & 0.967&\cite{Lin87} &   j   \\
\dec 276.0-- &                           &
                           &     &       &      &             &       \\
\dec --922.0 &   21  $\sigma_{\rm ce}$   & \dec   32.0
                         &5--10\%& 1.138 &      &\cite{Als75} &   m   \\
\dec 276.9   &   20 d$\sigma_{\rm el}$   & \dec   18.9
                           &  4\%& 1.042 &      &\cite{Spe70} &   o   \\
\dec 287.0   &   54 d$\sigma_{\rm el}$   &          .
                           &  5\%&  .    & all
                                         &\cite{Bru86a,Bru91a}&j, l, o\\
\dec 287.0   &   14 d$\sigma_{\rm ce}$   & \dec   24.0
                           &  5\%& 1.201 & 0.985
                                                &\cite{Bru86b}&       \\
\dec 310.4   &   20 d$\sigma_{\rm el}$   & \dec   30.1
                           &  4\%& 1.039 &      &\cite{Spe70} &   o   \\
%\dec 335.0-- &                           &
%                           &     &       &      &             &      \\
%\dec --9aa.a &   3a  $\sigma_{\rm tot}$  &         .
%                           &1 mb &  .    &      &\cite{Sum82} &   m  \\
\dec 340.9   &   20 d$\sigma_{\rm el}$   & \dec   32.2
                           &  4\%& 1.044 & $-$0.850
                                                &\cite{Spe70} &   o   \\
\dec 348.7   &   38 d$\sigma_{\rm el}$   & \dec   42.4
                           &  4\%& 0.993 &      &\cite{Con68} & i, o  \\
\dec 355.0-- &                           &
                           &     &       &      &             &       \\
\dec --923.0 &   36  $\sigma_{\rm tot}$  &          .
                           &1.5\%&  .    & all  &\cite{Ham80b}& e, m  \\
\dec 353.3   &  119 d$\sigma_{\rm el}$   & \dec  116.4
                           &  5\%& 1.037 & 0.366&\cite{Cre83} & j, o  \\
\dec 369.1   &   19 d$\sigma_{\rm el}$   & \dec   15.7
                           &  4\%& 1.020 & 0.550
                                                &\cite{Spe70} & i, o  \\
\dec 374.0   &   39 d$\sigma_{\rm el}$   & \dec   24.3
                           &  5\%& 1.067 &      &\cite{Sak82} &   o   \\
\dec 388.0-- &                           &
                           &     &       &      &             &       \\
\dec --598.6 &   29  $\sigma_{\rm tot}$  & \dec   35.4
                           &0.7\%& 0.973 &      &\cite{Clo84} &       \\
\dec 392.4   &   19 d$\sigma_{\rm el}$   &          .
                           &  5\%&  .    & all  &\cite{Kag87} &  l    \\
\dec 392.4   &   15 d$\sigma_{\rm ce}$   &          .
                           &  5\%&  .    & all  &\cite{Nak84b}&   f   \\
%\dec 395.9-- &                           &
%                           &     &       &      &             &       \\
%\dec --737.4 &   28  $\sigma_{\rm tot}$  &          .
%                     &1.9--2.1 mb&  .    &      &\cite{Kam80,Nak84a}& \\
\dec 404.3   &   40 d$\sigma_{\rm el}$   & \dec   38.6
                           &  4\%& 0.994 & $-$0.575, $-$0.925
                                                &\cite{Con68} & i, o  \\
\dec 406.0-- &                           &
                           &     &       &      &             &       \\
\dec --922.0 &   30 d$\sigma_{\rm el}$   & \dec   36.7
                           &  4\%& 0.896 &      &\cite{Als79} &   n   \\
\end{tabular}
\label{tab:data}
\end{table}
\addtocounter{table}{-1}

\widetext
\begin{table}
\caption{({\it Continued}.)}
\begin{tabular}{cccccccc}
$p_{\rm lab}$& $\rm No.^a$ &
              &  Norm  &  Pred.    &          &           &          \\
(MeV/c) &  $\rm type^b$ & $\chi^2_{\rm min}$ &
 error & $\rm norm^c$ & $\rm Rejected^d$ & Ref. & Comm.   \\
\tableline
\dec 406.0   &  119 d$\sigma_{\rm el}$   & \dec   99.8
                           &  5\%& 1.029 & 0.990, 0.750, 0.578
                                                &\cite{Cre83} & j, o  \\
\dec 411.2   &   38 d$\sigma_{\rm el}$   & \dec   37.9
                           &  5\%& 1.025 & $-$0.925
                                                &\cite{Sak82} & i, o  \\
\dec 413.4   &    7 d$\sigma_{\rm el}$   & \dec    5.6
                           &  5\%& 1.054 & 0.992&\cite{Iwa81} & j, o  \\
\dec 424.5   &    7 d$\sigma_{\rm el}$   &          .
                           &  5\%&  .    & all  &\cite{Iwa81} &e, j, o \\
\dec 428.0   &   10 d$\sigma_{\rm ce}$   & \dec   12.3
                           & 20\%& 1.221 &      &\cite{Biz68} &       \\
\dec 435.8   &    7 d$\sigma_{\rm el}$   & \dec    1.6
                           &  5\%& 1.017 & 0.992&\cite{Iwa81} & j, o  \\
\dec 439.0   &   27 d$\sigma_{\rm el}$   &          .
                           & 10\%&  .    & all  &\cite{Per91} &   l   \\
\dec 439.0   &   24  $A_{y{\rm ,el}}$    & \dec   38.8
                           &  5\%& 1.068 & 0.851&\cite{Per91} &   o   \\
\dec 439.9   &   39 d$\sigma_{\rm el}$   & \dec   42.0
                           &  5\%& 1.031 &      &\cite{Sak82} &   o   \\
\dec 440.8   &   38 d$\sigma_{\rm el}$   & \dec   61.4
                           &  5\%& 1.035 &      &\cite{Sak82} & i, o  \\
\dec 444.1   &   38 d$\sigma_{\rm el}$   & \dec   35.0
                           &  4\%& 0.972 & 0.175, $-$0.825, $-$0.875
                                                &\cite{Con68} & i, o  \\
\dec 446.0   &  119 d$\sigma_{\rm el}$   & \dec  115.3
                           &  5\%& 1.021 &      &\cite{Cre83} & j, o  \\
\dec 447.1   &    7 d$\sigma_{\rm el}$   & \dec    6.9
                           &  5\%& 1.050 & 0.992&\cite{Iwa81} & j, o  \\
\dec 458.3   &    8 d$\sigma_{\rm el}$   & \dec    2.0
                           &  5\%& 0.994 & 0.996&\cite{Iwa81} & j, o  \\
\dec 467.5   &   39 d$\sigma_{\rm el}$   & \dec   32.7
                           &  4\%& 1.039 & $-$0.925
                                                &\cite{Con68} & i, o  \\
\dec 467.8   &   39 d$\sigma_{\rm el}$   & \dec   24.0
                           &  5\%& 1.056 &      &\cite{Sak82} &   o   \\
\dec 469.2   &    8 d$\sigma_{\rm el}$   & \dec    8.2
                           &  5\%& 1.013 & 0.996&\cite{Iwa81} & j, o  \\
\dec 479.3   &  119 d$\sigma_{\rm el}$   & \dec  109.3
                           &  5\%& 1.003 & 0.919, 0.873, 0.697
                                                &\cite{Cre83} & j, o  \\
\dec 480.0   &   10 d$\sigma_{\rm ce}$   & \dec   14.3
                        &$\infty$& 1.154 &      &\cite{Tsu83} &   g   \\
\dec 481.2   &    8 d$\sigma_{\rm el}$   & \dec    7.2
                           &  5\%& 1.048 & 0.996&\cite{Iwa81} & j, o  \\
\dec 490.1   &   37 d$\sigma_{\rm el}$   &          .
                           &  5\%&  .    & all  &\cite{Kag87} &  l    \\
\dec 490.1   &   15 d$\sigma_{\rm ce}$   & \dec   12.5
                           &  5\%& 1.068 & $-$0.193
                                                &\cite{Nak84b}&       \\
\dec 490.6   &   39 d$\sigma_{\rm el}$   & \dec   47.5
                           &  5\%& 0.983 &      &\cite{Sak82} &   o   \\
\dec 492.7   &    8 d$\sigma_{\rm el}$   & \dec    4.2
                           &  5\%& 1.014 & 0.996&\cite{Iwa81} & j, o  \\
\dec 497.0   &   14  $A_{y{\rm ,el}}$    & \dec    7.3
                        &$\infty$& 0.718 &      &\cite{Kun88,Kun89}&h \\
\dec 498.7   &   37 d$\sigma_{\rm el}$   & \dec   27.7
                           &  4\%& 1.004 &      &\cite{Con68} & i, o  \\
\dec 503.8   &    8 d$\sigma_{\rm el}$   & \dec   13.3
                           &  5\%& 1.047 & 0.996&\cite{Iwa81} & j, o  \\
\dec 504.7   &   39 d$\sigma_{\rm el}$   & \dec   14.3
                           &  5\%& 1.021 &      &\cite{Sak82} &   o   \\
\dec 505.0   &   54 d$\sigma_{\rm el}$   &          .
                           &  5\%&  .    & all
                                         &\cite{Bru86a,Bru91a}&j, l, o\\
\dec 505.0   &   14 d$\sigma_{\rm ce}$   & \dec   30.1
                           &  5\%& 1.034 & 0.574&\cite{Bru86b}&       \\
\dec 508.0   &  119 d$\sigma_{\rm el}$   & \dec  105.3
                           &  5\%& 1.019 & 0.663, 0.530
                                                &\cite{Cre83} & j, o  \\
\dec 508.9   &   39 d$\sigma_{\rm el}$   & \dec   28.7
                           &  5\%& 1.025 &      &\cite{Sak82} &   o   \\
\dec 516.0   &    8 d$\sigma_{\rm el}$   & \dec    5.5
                           &  5\%& 1.018 & 0.996&\cite{Iwa81} & j, o  \\
\dec 523.0   &   15  $A_{y{\rm ,el}}$    & \dec    8.3
                        &$\infty$& 0.786 &      &\cite{Kun88,Kun89}&h \\
\dec 524.8   &   36 d$\sigma_{\rm el}$   & \dec   32.2
                           &  4\%& 1.020 &      &\cite{Con68} & i, o  \\
\dec 525.9   &   39 d$\sigma_{\rm el}$   & \dec   45.0
                           &  5\%& 1.053 &      &\cite{Sak82} &   o   \\
\end{tabular}
\end{table}
\addtocounter{table}{-1}

\widetext
\begin{table}
\caption{({\it Continued}.)}
\begin{tabular}{cccccccc}
$p_{\rm lab}$& $\rm No.^a$ &
              &  Norm  &  Pred.    &          &           &          \\
(MeV/c) &  $\rm type^b$ & $\chi^2_{\rm min}$ &
 error & $\rm norm^c$ & $\rm Rejected^d$ & Ref. & Comm.   \\
\tableline
\dec 528.2   &    8 d$\sigma_{\rm el}$   & \dec    3.2
                           &  5\%& 1.005 & 0.996&\cite{Iwa81} & j, o  \\
\dec 533.6   &  119 d$\sigma_{\rm el}$   & \dec  133.7
                           &  5\%& 1.029 &      &\cite{Cre83} & j, o  \\
\dec 537.0   &   10 d$\sigma_{\rm ce}$   & \dec   19.4
                        &$\infty$& 1.199 &      &\cite{Tsu83} &   g   \\
\dec 540.6   &    8 d$\sigma_{\rm el}$   & \dec   10.9
                           &  5\%& 1.015 & 0.996&\cite{Iwa81} & j, o  \\
\dec 543.2   &   39 d$\sigma_{\rm el}$   & \dec   38.4
                           &  5\%& 1.065 & $-$0.975
                                                &\cite{Sak82} &   o   \\
\dec 544.0   &   33 d$\sigma_{\rm el}$   &          .
                           & 10\%&  .    & all  &\cite{Per91} &   l   \\
\dec 544.0   &   30  $A_{y{\rm ,el}}$    & \dec   37.5
                           &  5\%& 1.046 & $\geq$0.883, \#=3
                                                &\cite{Per91} &   o   \\
\dec 546.0   &   23  $A_{y{\rm ,ce}}$    & \dec   36.1
                           &  4\%& 0.959 &      &\cite{Bir91} &       \\
\dec 549.4   &   10 d$\sigma_{\rm ce}$   & \dec    7.1
                           & 20\%& 1.258 &      &\cite{Biz68} &       \\
\dec 550.0   &   67 d$\sigma_{\rm el}$   & \dec   76.0
                           &  5\%& 1.006 & $\geq$0.995, \#=3; &&      \\
                      &&&&& 0.910, 0.883, 0.869 &\cite{Sch89} &   j   \\
\dec 553.1   &   34 d$\sigma_{\rm el}$   & \dec   38.6
                           &  4\%& 0.981 &      &\cite{Con68} & i, o  \\
\dec 553.4   &    8 d$\sigma_{\rm el}$   & \dec    2.4
                           &  5\%& 1.017 & 0.996, 0.972
                                                &\cite{Iwa81} & j, o  \\
\dec 556.9   &  119 d$\sigma_{\rm el}$   & \dec  125.4
                           &  5\%& 1.025 & 0.908&\cite{Cre83} & j, o  \\
\dec 558.5   &   39 d$\sigma_{\rm el}$   & \dec   45.4
                           &  5\%& 1.040 &      &\cite{Sak82} &   o   \\
\dec 565.5   &    8 d$\sigma_{\rm el}$   & \dec    5.7
                           &  5\%& 1.006 & 0.996&\cite{Iwa81} & j, o  \\
\dec 568.4   &   37 d$\sigma_{\rm el}$   & \dec   34.3
                           &  5\%& 1.040 & $-$0.675, $-$0.825
                                                &\cite{Sak82} & i, o  \\
\dec 577.2   &   36 d$\sigma_{\rm el}$   & \dec   36.0
                           &  4\%& 0.983 &      &\cite{Con68} & i, o  \\
\dec 578.1   &    9 d$\sigma_{\rm el}$   & \dec    6.2
                           &  5\%& 1.014 & 0.999&\cite{Iwa81} & j, o  \\
\dec 578.3   &  119 d$\sigma_{\rm el}$   & \dec  132.3
                           &  5\%& 1.047 &      &\cite{Cre83} & j, o  \\
\dec 584.0   &   10 d$\sigma_{\rm ce}$   & \dec   15.7
                        &$\infty$& 1.112 &      &\cite{Tsu83} &   g   \\
\dec 590.0   &   39 d$\sigma_{\rm el}$   &          .
                           &  5\%&  .    & all
                                         &\cite{Bru86a,Bru91a}&j, l, o\\
\dec 590.0   &   15 d$\sigma_{\rm ce}$   & \dec   32.8
                           &  5\%& 1.092 & 0.996, $-$0.574
                                                &\cite{Bru86b}&       \\
\dec 591.2   &    9 d$\sigma_{\rm el}$   & \dec    6.5
                           &  5\%& 1.029 & 0.999&\cite{Iwa81} & j, o  \\
\dec 591.2   &   39 d$\sigma_{\rm el}$   &          .
                           &  5\%&  .    & all  &\cite{Kag87} &  l    \\
\dec 591.2   &   15 d$\sigma_{\rm ce}$   & \dec   18.0
                           &  5\%& 1.058 & $-$0.358
                                                &\cite{Nak84b}&       \\
\dec 596.5   &   38 d$\sigma_{\rm el}$   & \dec   49.5
                           &  5\%& 1.075 &      &\cite{Sak82} &   o   \\
\dec 599.2   &   33 d$\sigma_{\rm el}$   & \dec   15.8
                           &  4\%& 0.997 &      &\cite{Con68} & i, o  \\
\dec 604.0   &    9 d$\sigma_{\rm el}$   & \dec    7.6
                           &  5\%& 0.987 & 0.998&\cite{Iwa81} & j, o  \\
\dec 615.0   &   38 d$\sigma_{\rm el}$   & \dec   48.1
                           &  5\%& 1.046 & $-$0.575
                                                &\cite{Sak82} &   o   \\
\dec 617.0   &    9 d$\sigma_{\rm el}$   & \dec    6.7
                           &  5\%& 0.953 & 0.998&\cite{Iwa81} & j, o  \\
\dec 630.0   &   10 d$\sigma_{\rm ce}$   & \dec    9.3
                        &$\infty$& 1.073 &      &\cite{Tsu83} &   g   \\
\dec 630.9   &    9 d$\sigma_{\rm el}$   & \dec    4.6
                           &  5\%& 0.997 & 0.998&\cite{Iwa81} & j, o  \\
\dec 639.6   &   38 d$\sigma_{\rm el}$   & \dec   17.1
                           &  5\%& 0.995 & $-$0.175
                                                &\cite{Sak82} &   o   \\
\dec 644.7   &    9 d$\sigma_{\rm el}$   & \dec    7.8
                           &  5\%& 0.996 & 0.998&\cite{Iwa81} & j, o  \\
\dec 656.0   &   21  $A_{y{\rm ,ce}}$    & \dec   21.9
                           &  4\%& 0.963 &      &\cite{Bir91,Bir90}&  \\
\dec 658.1   &   38 d$\sigma_{\rm el}$   & \dec   37.4
                           &  5\%& 0.972 & 0.225, $-$0.675, $-$0.975
                                                &\cite{Sak82} &   o   \\
\end{tabular}
\end{table}
\addtocounter{table}{-1}

\widetext
\begin{table}
\caption{({\it Continued}.)}
\begin{tabular}{cccccccc}
$p_{\rm lab}$& $\rm No.^a$ &
              &  Norm  &  Pred.    &          &           &          \\
(MeV/c) &  $\rm type^b$ & $\chi^2_{\rm min}$ &
 error & $\rm norm^c$ & $\rm Rejected^d$ & Ref. & Comm.   \\
\tableline
\dec 658.6   &    9 d$\sigma_{\rm el}$   & \dec    8.9
                           &  5\%& 1.004 & 0.998&\cite{Iwa81} & j, o  \\
\dec 670.0   &   10 d$\sigma_{\rm ce}$   & \dec    5.5
                        &$\infty$& 1.165 &      &\cite{Tsu83} &   g   \\
\dec 671.5   &    9 d$\sigma_{\rm el}$   & \dec    3.3
                           &  5\%& 0.987 & 0.998&\cite{Iwa81} & j, o  \\
\dec 679.0   &   26 d$\sigma_{\rm el}$   &          .
                        &$\infty$&  .    & all  &\cite{Kun89} &  l    \\
\dec 679.0   &   27  $A_{y{\rm ,el}}$    & \dec   23.1
                        &$\infty$& 0.846 & 0.540&\cite{Kun88,Kun89}&h \\
\dec 679.0   &    1  $D_{yy}$            & \dec    1.4
                           & --  &  .    &      &\cite{Kun91} & p, q  \\
\dec 679.1   &    4  $A_{y{\rm ,el}}$    & \dec    6.3
                           &  5\%& 0.983 &      &\cite{Ohs73} &   o   \\
\dec 680.1   &   38 d$\sigma_{\rm el}$   & \dec   40.2
                           &  5\%& 1.003 &      &\cite{Sak82} &   o   \\
\dec 686.1   &    9 d$\sigma_{\rm el}$   & \dec    3.9
                           &  5\%& 0.986 & 0.998&\cite{Iwa81} & j, o  \\
\dec 689.0   &   39 d$\sigma_{\rm el}$   &          .
                           &  5\%&  .    & all  &\cite{Kag87} &  l    \\
\dec 689.0   &   16 d$\sigma_{\rm ce}$   & \dec   26.5
                           &  5\%& 1.010 & $-$0.139
                                                &\cite{Nak84b}&       \\
\dec 690.0   &   89 d$\sigma_{\rm el}$   & \dec  103.5
                           &  4\%& 0.991 & 0.370&\cite{Eis76} &       \\
\dec 693.0   &   34 d$\sigma_{\rm ce}$   & \dec   39.3
                           & 10\%& 1.069 & $-$0.075
                                                &\cite{Bir90} &  r    \\
\dec 696.1   &   21 d$\sigma_{\rm el}$   & \dec   16.4
                           &  4\%& 1.026 &      &\cite{Koh72} &       \\
\dec 696.1   &   16 d$\sigma_{\rm ce}$   & \dec   21.0
                           &  4\%& 1.050 &      &\cite{Koh72} &       \\
\dec 697.0   &   24 d$\sigma_{\rm el}$   &          .
                           & 10\%&  .    & all  &\cite{Ber89} &   l   \\
\dec 697.0   &   33  $A_{y{\rm ,el}}$    & \dec   20.3
                           &  5\%& 1.022 & 0.629&\cite{Ber89} &   o   \\
\dec 698.0   &   10 d$\sigma_{\rm ce}$   & \dec    7.1
                        &$\infty$& 1.237 &      &\cite{Tsu83} &   g   \\
\dec 700.0   &    4  $A_{y{\rm ,el}}$    & \dec    2.1
                           &  5\%& 0.991 &      &\cite{Kim82} &   o   \\
\dec 701.1   &    9 d$\sigma_{\rm el}$   & \dec    3.5
                           &  5\%& 1.000 & 0.998&\cite{Iwa81} & j, o  \\
\dec 715.3   &    9 d$\sigma_{\rm el}$   & \dec   10.6
                           &  5\%& 1.002 & 0.998&\cite{Iwa81} & j, o  \\
\dec 728.0   &   10 d$\sigma_{\rm ce}$   & \dec    2.9
                        &$\infty$& 1.105 &      &\cite{Tsu83} &   g   \\
\dec 757.0   &   72 d$\sigma_{\rm el}$   & \dec   95.7
                           &  5\%& 1.055 & $\geq$0.996, \#=3
                                                &\cite{Sch89} &   j   \\
\dec 767.0   &   22  $A_{y{\rm ,ce}}$    & \dec   25.5
                           &  4\%& 1.113 &      &\cite{Bir91} &       \\
\dec 780.5   &   39 d$\sigma_{\rm el}$   &          .
                           &  5\%&  .    & all  &\cite{Kag87} &  l    \\
\dec 780.5   &   15 d$\sigma_{\rm ce}$   & \dec   14.0
                           &  5\%& 0.974 &      &\cite{Nak84b}&       \\
\dec 783.0   &   30 d$\sigma_{\rm el}$   &          .
                        &$\infty$&  .    & all  &\cite{Kun89} &  l    \\
\dec 783.0   &   30  $A_{y{\rm ,el}}$    & \dec   30.1
                           &4.5\%& 0.944 & $-$0.300
                                                &\cite{Kun88,Kun89}&  \\
\dec 783.0   &    3  $D_{yy}$            & \dec    4.9
                           & --  &  .    &      &\cite{Kun91} & p, q  \\
\dec 790.0   &   95 d$\sigma_{\rm el}$   & \dec  101.5
                           &  4\%& 1.034 &      &\cite{Eis76} &       \\
\dec 860.0   &   95 d$\sigma_{\rm el}$   & \dec   70.5
                           &  4\%& 1.045 & 0.510&\cite{Eis76} &       \\
\dec 875.0   &   23  $A_{y{\rm ,ce}}$    & \dec   20.0
                           &  4\%& 0.995 &      &\cite{Bir91} &       \\
\dec 886.0   &   34 d$\sigma_{\rm el}$   &          .
                        &$\infty$&  .    & all  &\cite{Kun89} &  l    \\
\dec 886.0   &   34  $A_{y{\rm ,el}}$    & \dec   38.5
                           &4.5\%& 0.992 &      &\cite{Kun88,Kun89}&  \\
\dec 886.0   &    1  $D_{yy}$            & \dec    1.2
                           & --  &  .    &      &\cite{Kun91} & p, q  \\
\dec 910.0   &   19 d$\sigma_{\rm el}$   &          .
                        &$\infty$&  .    & all  &\cite{Alb72} &  f, g \\
\dec 910.0   &   21  $A_{y{\rm ,el}}$    & \dec   14.7
                           &  5\%& 0.989 &      &\cite{Alb72} &       \\
\end{tabular}
\end{table}

\begin{itemize}
\item[a] The number includes all published data, except those given
         as 0.0$\pm$0.0 (see Comment i), and those having
         $p_{\mbox{\scriptsize lab}}>925$ MeV/c (see Comment m).
\item[b] The subscripts ``el'' and ``ce'' denote observables
         in the elastic $\overline{p}p \rightarrow \overline{p}p$
         and charge-exchange $\overline{p}p \rightarrow \overline{n}n$
         reactions, respectively. ``d$\sigma$'' denotes
         a differential cross section d$\sigma$/d$\Omega$,
         $A_y$ a polarization-type datum (asymmetry or
         analyzing-power), and $D_{yy}$ a depolarization
         datum. $\sigma_{\rm tot}$ stands for total cross section,
         $\sigma_{\rm ann}$ for (total) annihilation cross section,
         and $\sigma_{\rm ce}$ for integrated charge-exchange
         cross section.
\item[c] Normalization, predicted by the analysis, with which the
         experimental values should be multiplied before
         comparison with the theoretical values.
\item[d] Tabulated is $p_{\mbox{\scriptsize lab}}$ in MeV/c
         or $\cos\theta$. The notation ``$\geq$0.925, \#=6''
         e.g. means that the 6 points with $\cos\theta\geq$0.925
         are rejected.
\item[e]  Group rejected due to improbable low $\chi^2_{\rm min}$.
\item[f]  Group rejected due to improbable high $\chi^2_{\rm min}$.
\item[g]  Floated normalization. Data are relative only.
\item[h]  Normalization floated by us, since the norm contributes
          much more than 9 to $\chi^2_{\rm min}$.
\item[i]  Data points given as 0.0$\pm$0.0 not included.
\item[j]  Coulomb-nuclear--interference measurement. Data points
          in the extreme forward angular region are rejected
          when they contain multiple-scattering effects.
\item[k]  Data points at low momenta rejected (see text).
\item[l]  Problematic differential cross section. Not included
          in the database. See the text, Sect. VIIIB
          and Tables II and III.
\item[m]  Part of a group of data with points having
          $p_{\mbox{\scriptsize lab}}>925$ MeV/c.
\item[n]  Elastic differential cross sections as a function of
          momentum taken at backward angle $\cos\theta=-0.994$.
\item[o]  Normalization error assumed by us, since no
          clear number is stated in the reference.
\item[p]  Depolarization data. Not included in the fit, in view
          of large error bars.
\item[q]  Normalization error taken to be zero, in view
          of large error bars of these data.
\item[r]  Data points taken at the same angles averaged.
\end{itemize}

\newpage
\begin{table}
\caption{Corrections applied to elastic differential
         cross sections from Eisenhandler {\it et al}., from PS172,
         PS173, and PS198 from LEAR, and from KEK. The number
         of data includes all published points. The column labeled
         ``Syst. error'' gives the approximate point-to-point error
         that has to be added quadratically to the statistical error
         to reach $\chi^2_{\rm min} \approx N_{\rm df}$.
         The column labeled ``$\chi^2_{\rm unc}$'' gives the result
         of the Legendre fit without any corrections, except that
         PS173 points at forward angles contaminated by
         multiple-scattering effects are removed. For the
         values of $\chi^2_{\rm min}$ after the corrections,
         see Table~\protect\ref{tab:Legendre}.}
\begin{tabular}{clclcrcc}
$p_{\rm lab}$ &       & No.  &  & Syst. &  &  &  \\
  (MeV/c)     & Group & data & Rejected points ($\cos\theta$) &
 error & $\chi^2_{\rm unc}$ & Ref. & Comm. \\ \tableline
 181.0 & PS173  & 46  & $>0.92$, \#=6
                            &     &  41.0 & \cite{Bru86a,Bru91a} & a  \\
 287.0 & PS173  & 54  & $>0.95$, \#=9; 0.345, 0.199, 0.101, 0.051, $-$0.345
                            & 5\% & 160.7 & \cite{Bru86a,Bru91a} & a  \\
 392.4 & KEK    & 19  &     & 2\% &  19.2 & \cite{Kag87}         & b  \\
 439.0 & PS198  & 27  &     & 4\% &  28.0 & \cite{Per91}         &    \\
 490.1 & KEK    & 18  &     & 4\% & 202.0 &                      & b  \\
       &        & 19  &     &     &  15.0 & \cite{Kag87}         & c  \\
 505.0 & PS173  & 54  & $>0.98$, \#=3; 0.96, 0.67
                            &     &  59.5 & \cite{Bru86a,Bru91a} & a  \\
 544.0 & PS198  & 33  &     &     &  23.1 & \cite{Per91}         &    \\
 590.0 & PS173  & 39  & $>0.99$, \#=2
                            & 3\% &  39.0 & \cite{Bru86a,Bru91a} & a  \\
 591.2 & KEK    & 19  & 0.825
                            & 3\% & 276.2 &                      & b  \\
       &        & 20  & 0.275, $-$0.075
                            & 2\% &  36.0 & \cite{Kag87}         & c  \\
 679.0 & PS172  & 26  & 0.50, $-$0.50
                            &     &  82.7 & \cite{Kun89}         & d  \\
 689.0 & KEK    & 15  &     & 3\% &  90.5 &                      & b  \\
       &        & 24  &     & 3\% &  24.2 & \cite{Kag87}         & c  \\
 690.0 & Eisenh.& 89  & 0.37&     & 101.9 & \cite{Eis76}         &    \\
 697.0 & PS198  & 24  &     & 6\% & 174.0 & \cite{Ber89}         &    \\
 780.5 & KEK    & 14  &     & 2\% &  30.5 &                      & b  \\
       &        & 25  &     &     &  16.4 & \cite{Kag87}         & c  \\
 783.0 & PS172  & 30  & 0.58, $-$0.58
                            &     &  31.9 & \cite{Kun89}         & d  \\
 790.0 & Eisenh.& 95  &     &     &  88.2 & \cite{Eis76}         &    \\
 860.0 & Eisenh.& 95  &     &     &  69.4 & \cite{Eis76}         &    \\
 886.0 & PS172  & 34  & 0.66, 0.54, 0.46, $-$0.18, $-$0.66
                            &     &  77.6 & \cite{Kun89}         & d  \\
\end{tabular}
\label{tab:flaws}
\end{table}

\begin{itemize}
\item[a]  Data points in the extreme forward angular region are
          rejected because they contain multiple-scattering effects.
\item[b]  Data in the ``one-prong'' region. For a discussion, see the text.
\item[c]  Data in the ``two-prong'' region. For a discussion, see the text.
\item[d]  The most forward and the most backward data point are rejected
          (F. Bradamante, private communication).
\end{itemize}

\newpage
\begin{table}
\caption{Legendre-polynomial fits to elastic differential cross sections
         from Eisenhandler {\it et al}., PS172, PS173, and PS198 from LEAR,
         and from KEK. The corrections summarized in
         Table~\protect\ref{tab:flaws} are taken into account. The number
         of data is the number of published data minus the rejected points.
         For the KEK groups at 490, 591, 689, and 781 MeV/c the first
         line gives the results for the ``one-prong'' region, and the
         second line the results for the ``two-prong'' region.}
\begin{tabular}{clc|ccccccc|r}
$p_{\rm lab}$ && No. &&&&&&& \\
  (MeV/c) &  Group  & data & $a_0$ & $a_1/a_0$ &
 $a_2/a_0$ & $a_3/a_0$ & $a_4/a_0$ & $a_5/a_0$ & $a_6/a_0$ &
 $\chi^2_{\rm min}$  \\ \tableline
 181.0 & PS173  & 40  &
 7.02(25) & 0.92(06) & 0.46(07) & & & & & 41.0 \\
 287.0 & PS173  & 40  &
 5.10(07) & 1.58(03) & 0.82(03) & 0.15(02) & & & & 36.0 \\
 392.4 & KEK    & 19  &
 5.49(06) & 1.92(04) & 1.35(05) & 0.35(03) & & & & 15.6 \\
 439.0 & PS198  & 27  &
 4.09(08) & 1.93(06) & 1.63(07) & 0.67(08) &
 0.39(09) & 0.19(07) & 0.09(04) & 19.6 \\
 490.1 & KEK    & 18  &
 5.07(10) & 1.95(05) & 1.55(07) & 0.50(05) &
 0.10(03) & & & 15.4 \\
       &        & 19  &
 5.61(45) & 1.91(21) & 1.89(28) & 0.72(13) &
 0.38(10) & & & 15.0 \\
 505.0 & PS173  & 49  &
 4.18(05) & 1.90(04) & 1.76(04) & 0.96(05) &
 0.49(06) & 0.11(05) & 0.06(04) & 36.1 \\
 544.0 & PS198  & 33  &
 3.79(05) & 2.08(05) & 2.16(06) & 1.16(06) &
 0.57(06) & 0.12(04) & 0.04(02) & 23.1 \\
 590.0 & PS173  & 37  &
 3.73(06) & 2.11(05) & 2.09(05) & 1.25(06) &
 0.66(06) & 0.26(05) & 0.10(03) & 32.2 \\
 591.2 & KEK    & 18  &
 4.42(08) & 2.12(05) & 1.95(06) & 0.83(04) &
 0.16(03) & & & 13.8 \\
       &        & 18  &
 5.24(33) & 2.07(18) & 2.08(22) & 0.99(11) &
 0.38(07) & & & 15.2 \\
 679.0 & PS172  & 24  &
 1.96(02) & 2.59(05) & 1.88(02) & 1.39(04) & & & & 18.8 \\
 689.0 & KEK    & 15  &
 3.52(22) & 2.27(15) & 2.35(18) & 1.46(11) &
 0.63(14) & 0.17(07) & & 9.8 \\
       &        & 24  &
 3.21(15) & 2.16(14) & 2.14(17) & 1.20(09) &
 0.38(06) & & & 19.0 \\
 690.0 & Eisenh. & 88  &
 4.28(07) & 2.28(06) & 2.45(08) & 1.67(07) &
 0.76(06) & 0.23(04) & 0.03(02) & 81.6 \\
 697.0 & PS198  & 24  &
 3.54(09) & 2.22(08) & 2.30(09) & 1.43(08) &
 0.51(05) & 0.10(02) & & 21.5 \\
 780.5 & KEK    & 14  &
 3.14(23) & 2.37(17) & 2.72(23) & 2.10(16) &
 1.07(16) & 0.29(07) & & 9.1 \\
       &        & 25  &
 2.39(10) & 2.07(11) & 2.39(16) & 1.36(08) &
 0.66(07) & & & 16.4 \\
 783.0 & PS172  & 28  &
 2.55(07) & 2.73(17) & 2.45(11) & 2.26(20) &
 0.67(05) & 0.38(07) & & 29.8 \\
 790.0 & Eisenh. & 95  &
 3.87(08) & 2.36(07) & 2.69(09) & 2.08(09) &
 1.09(07) & 0.42(05) & 0.12(03) & 88.2 \\
 860.0 & Eisenh. & 95  &
 3.67(06) & 2.41(06) & 2.83(07) & 2.33(07) &
 1.34(05) & 0.56(04) & 0.17(02) & 69.4 \\
 886.0 & PS172  & 29  &
 2.44(04) & 2.66(08) & 2.64(06) & 2.46(09) &
 0.95(03) & 0.47(04) & & 22.2 \\
\end{tabular}
\label{tab:Legendre}
\end{table}

\newpage
\begin{table}
\caption{$P$-matrix parameters of the different partial waves.
         $V_0$ and $V_1$ are the real parts of the short-range
         spherical-well potential, for isospin $I=0$ and $I=1$
         respectively. $W$ is the isospin-independent imaginary part.
         The mixing angles that take care of the short-range $I=0$
         tensor force are: $\theta_1$ = $55.7^{\circ}\pm1.3^{\circ}$
         for the $^3C_1$ state,
         $\theta_2$ = $45.8^{\circ}\pm1.1^{\circ}$
         for the $^3C_2$ state, and
         $\theta_2$ = $10.7^{\circ}\pm4.7^{\circ}$
         for the $^3C_3$ state.
         The quoted errors are defined as the change in each
         parameter that gives a maximal rise in $\chi^2_{\rm min}$
         of 1 when the remaining parameters are refitted.}
\begin{tabular}{c|ccc}
partial wave  & $V_0$ (MeV) & $V_1$ (MeV) & $W$ (MeV)   \\
 \tableline
 $^1S_0$     &      0     &        0      &  $-99(6)$   \\
 $^3S_1$     & $-151(6)$  &    $-17(3)$   & $-100(3)$   \\
 $^1P_1$     &      0     &        0      &  $-90(7)$   \\
 $^3P_0$     & $-132(9)$  &    $178(19)$  & $-156(9)$   \\
 $^3P_1$     &  $155(12)$ &    $-66(3)$   &  $-97(4)$   \\
 $^3P_2$     & $-136(4)$  &    $-69(3)$   & $-142(3)$   \\
 $^1D_2$     &      0     &        0      & $-105(12)$  \\
 $^3D_1$     &  $-215(7)$ &     $33(16)$  & $-106(5)$   \\
 $^3D_2$     &  $-38(12)$ &   $-198(4)$   & $-110(4)$   \\
 $^3D_3$     & $-152(6)$  &   $-102(5)$   & $-163(4)$   \\
 $^3F_2$     & $-101(20)$ &   $-250(14)$  & $-179(7)$   \\
\end{tabular}
\label{tab:pars}
\end{table}

\newpage
\begin{table}
\caption{Partial elastic and charge-exchange cross sections in mb.}
\begin{tabular}{c|cccc|cccc}
 &\multicolumn{4}{c|}{$\overline{p}p\rightarrow\overline{p}p$}
 &\multicolumn{4}{c}{$\overline{p}p\rightarrow\overline{n}n$} \\
 $p_{\mbox{\scriptsize lab}}$ (MeV/c)
                       &    200    &    400    &    600    &    800
                       &    200    &    400    &    600    &    800   \\
 \tableline
 $^1S_0$               & \dec 14.6 & \dec  6.8 & \dec  3.7 & \dec 1.9
                       & \dec  0.5 &           &           &          \\
 $^1P_1$               & \dec  1.7 & \dec  2.6 & \dec  2.5 & \dec 2.4
                       & \dec  1.3 & \dec  0.4 &           &          \\
 $^1D_2$               & \dec  0.1 & \dec  0.4 & \dec  0.7 & \dec 0.8
                       & \dec  0.1 & \dec  0.4 & \dec  0.2 &          \\
 $^1F_3$               &           &           & \dec  0.1 & \dec 0.2
                       &           & \dec  0.1 & \dec  0.1 & \dec 0.1 \\
 $^1G_4$               &           &           &           & \dec 0.1
                       &           &           & \dec  0.1 & \dec 0.1 \\
 $^3P_1$               & \dec  1.8 & \dec  7.6 & \dec  7.6 & \dec 6.2
                       & \dec  6.7 & \dec  6.2 & \dec  2.8 & \dec 1.4 \\
 $^3D_2$               & \dec  0.1 & \dec  0.3 & \dec  2.2 & \dec 3.9
                       & \dec  0.3 & \dec  2.6 & \dec  3.2 & \dec 2.0 \\
 $^3F_3$               &           & \dec  0.1 & \dec  0.1 & \dec 0.2
                       &           & \dec  0.4 & \dec  1.0 & \dec 1.4 \\
 $^3G_4$               &           &           &           &
                       &           &           & \dec  0.3 & \dec 0.4 \\
 $^3P_0$               & \dec  4.7 & \dec  4.6 & \dec  3.4 & \dec 2.6
                       & \dec  2.1 & \dec  1.4 & \dec  0.7 & \dec 0.3 \\
 $^3S_1$               & \dec 71.1 & \dec 29.6 & \dec 14.4 & \dec 7.9
                       & \dec  2.0 & \dec  0.5 & \dec  0.3 & \dec 0.3 \\
$^3S_1\rightarrow$$^3D_1$&\dec 0.2 & \dec  0.1 &           &
                         &\dec 0.6 & \dec  0.5 & \dec  0.1 &          \\
$^3D_1\rightarrow$$^3S_1$&\dec 0.2 & \dec  0.1 &           &
                         &\dec 1.4 & \dec  0.7 & \dec  0.1 &          \\
 $^3D_1$               &           & \dec  0.3 & \dec  0.8 & \dec 1.3
                       & \dec  0.1 & \dec  0.3 & \dec  0.5 & \dec 0.4 \\
 $^3P_2$               & \dec  6.3 & \dec 16.1 & \dec 15.5 & \dec 12.9
                       & \dec  0.8 & \dec  1.2 & \dec  0.5 & \dec 0.3 \\
$^3P_2\rightarrow$$^3F_2$&\dec 0.1 & \dec  0.1 & \dec  0.2 & \dec 0.2
                         &\dec 0.1 & \dec  0.4 & \dec  0.4 & \dec 0.2 \\
$^3F_2\rightarrow$$^3P_2$&\dec 0.1 & \dec  0.1 & \dec  0.2 & \dec 0.2
                         &\dec 0.3 & \dec  0.6 & \dec  0.4 & \dec 0.3 \\
 $^3F_2$               &           &           & \dec  0.1 & \dec 0.3
                       &           &           & \dec  0.1 &          \\
 $^3D_3$               &           & \dec  1.2 & \dec  5.0 & \dec 7.1
                       &           & \dec  0.4 & \dec  1.0 & \dec 0.6 \\
$^3D_3\rightarrow$$^3G_3$&         & \dec  0.1 & \dec  0.1 & \dec 0.1
                         &         & \dec  0.2 & \dec  0.3 & \dec 0.2 \\
$^3G_3\rightarrow$$^3D_3$&         & \dec  0.1 & \dec  0.1 & \dec 0.1
                         &         & \dec  0.3 & \dec  0.4 & \dec 0.3 \\
 $^3G_3$               &           &           &           &
                       &           &           &           & \dec 0.1 \\
 $^3F_4$               &           &           & \dec  0.3 & \dec 1.2
                       &           &           & \dec  0.1 & \dec 0.1 \\
$^3F_4\rightarrow$$^3H_4$&         &           &           & \dec 0.1
                         &         &           & \dec  0.1 & \dec 0.2 \\
$^3H_4\rightarrow$$^3F_4$&         &           &           & \dec 0.1
                         &         & \dec 0.1  & \dec  0.2 & \dec 0.2 \\
 $^3H_4$               &           &           &           &
                       &           &           &           &          \\
  rest                 & \dec 0.1  & \dec 0.1  & \dec 0.1  & \dec 0.1
                       & \dec 0.1  & \dec 0.1  & \dec 0.1  & \dec 0.9 \\
\tableline
singlet                & \dec 16.4 & \dec 9.8  & \dec 7.0  & \dec 5.4
                       & \dec  2.0 & \dec 0.9  & \dec 0.4  & \dec 0.2 \\
triplet                & \dec 84.7 & \dec 60.5 & \dec 50.1 & \dec 44.5
                       & \dec 14.4 & \dec 15.9 & \dec 12.6 & \dec 9.6 \\
 total                & \dec 101.1 & \dec 70.3 & \dec 57.1 & \dec 49.9
                       & \dec 16.4 & \dec 16.8 & \dec 13.0 & \dec 9.8 \\
 \tableline \tableline
 &\multicolumn{4}{c|}{$\overline{p}p\rightarrow$ all}
 &\multicolumn{4}{c}{$\overline{p}p\rightarrow$ mesons} \\
 $p_{\mbox{\scriptsize lab}}$ (MeV/c)
                       &    200    &    400    &    600    &    800
                       &    200    &    400    &    600    &    800   \\
 \tableline
                       & \dec 314.7 & \dec 193.9 & \dec 151.8 & \dec 128.5
                       & \dec 197.2 & \dec 106.8 & \dec  81.7 & \dec  68.8 \\
\end{tabular}
\label{tab:partxs}
\end{table}


\begin{references}
\bibitem[*]{mail}   Present address: Theoretical Division, Mail Stop B283,
                    Los Alamos National Laboratory, Los Alamos NM 87545,
                    USA; electronic address: timmer@t5zia.LANL.GOV.
\bibitem[**]{email} Electronic address: U634999@HNYKUN11.BITNET.
\bibitem{San83}     W.A. van der Sanden, A.H. Emmen, and
                    J.J. de Swart, Report THEF-NYM-83.11,
                    Nijmegen, 1983 (unpublished).
\bibitem{Arn83}     R.A. Arndt, L.D. Roper, R.A. Bryan, R.B. Clark,
                    B.J. VerWest, and P. Signell,
                    Phys. Rev. D {\bf 28}, 97 (1983).
\bibitem{Arn87}     R.A. Arndt, J.S. Hyslop III, and L.D. Roper,
                    Phys. Rev. D {\bf 35}, 128 (1987).
\bibitem{Bys87}     J. Bystricky, C. Lechanoine-LeLuc, and
                    F. Lehar, J. Phys. (Paris) {\bf 48}, 199 (1987);
                    F. Lehar, C. Lechanoine-LeLuc, and
                    J. Bystricky, {\it ibid}. {\bf 48}, 199 (1987).
\bibitem{Ber88}     J.R. Bergervoet, P.C. van Campen,
                    W.A. van der Sanden,  and J.J. de Swart,
                    Phys. Rev. C {\bf 38}, 15 (1988).
\bibitem{Ber90}     J.R. Bergervoet, P.C. van Campen, R.A.M. Klomp,
                    J.-L. de Kok, T.A. Rijken, V.G.J. Stoks, and
                    J.J. de Swart,
                    Phys. Rev. C {\bf 41}, 1435 (1990).
\bibitem{Klo93}     V.G.J. Stoks, R.A.M. Klomp,
                    M.C.M. Rentmeester, and J.J. de Swart,
                    Phys. Rev. C {\bf 48}, 792 (1993).
\bibitem{Gra88}     I.L. Grach, B.O. Kerbikov, and Yu.A. Simonov,
                    Phys. Lett. B {\bf 208}, 309 (1988).
\bibitem{Mah88}     J. Mahalanabis, H.J. Pirner, and T.-A. Shibata,
                    Nucl.~Phys. {\bf A485}, 546 (1988).
\bibitem{Pir91}     H.J. Pirner, B. Kerbikov, and J. Mahalanabis,
                    Zeit. Phys. {\bf A338}, 111 (1991).
\bibitem{Fet77}     E. Fett, A. Haatuft, J.M. Olsen, W. Hart,
                    D. Waldren, P. Alibran, B. Degrange, R. Arnold,
                    J.P. Engel, and S. de Unamuno,
                    Nucl. Phys. {\bf B130}, 1 (1977).
\bibitem{Van88}     J. Vandermeulen,
                    Zeit. Phys. {\bf C37}, 563 (1988).
\bibitem{Loc92}     M.P. Locher and B.S. Zou,
                    Zeit. Phys. {\bf A341}, 191 (1992).
\bibitem{Bal58}     J.S. Ball and G.F. Chew,
                    Phys. Rev. {\bf 109}, 1385 (1958).
\bibitem{Spe67}     M.S. Spergel,
                    Il Nuovo Cimento A {\bf 47}, 410 (1967).
\bibitem{Dal77}     O.D. Dalkarov and F. Myhrer,
                    Il Nuovo Cimento {\bf 40A}, 152 (1977).
\bibitem{Del78}     A. Delville, P. Jasselette, and J. Vandermeulen,
                    Am. J. Phys. {\bf 46}, 907 (1978).
\bibitem{Bry68}     R.A. Bryan and R.J.N. Phillips,
                    Nucl. Phys. {\bf B5}, 201 (1968);
                    {\it ibid}. {\bf B7}, 481(E) (1968);
                    R.J.N. Phillips,
                    Rev. Mod. Phys. {\bf 39}, 681 (1967).
\bibitem{Myh77}     F. Myhrer and A. Gersten,
                    Il Nuovo Cimento {\bf 37A}, 21 (1977).
\bibitem{Dov80}     C.B. Dover and J.-M. Richard,
                    Phys. Rev. C {\bf 21}, 1466 (1980).
                    {\it ibid}. {\bf 25}, 1952 (1982).
\bibitem{Koh86}     M. Kohno and W. Weise,
                    Nucl. Phys. {\bf A454}, 429 (1986).
\bibitem{Shi87}     T.-A. Shibata,
                    Phys. Lett. B {\bf 189}, 232 (1987).
\bibitem{Hip89}     T. Hippchen, K. Holinde, and W. Plessas,
                    Phys. Rev. C {\bf 39}, 761 (1989).
\bibitem{Cot82}     J. C\^ot\'e, M. Lacombe, B. Loiseau,
                    B. Moussallam,  and R. Vinh Mau,
                    Phys. Rev. Lett. {\bf 48}, 1319 (1982);
                    M. Lacombe, B. Loiseau,
                    B. Moussallam,  and R. Vinh Mau,
                    Phys. Rev. C {\bf 29}, 1800 (1984).
\bibitem{Pig91}     M. Pignone, M. Lacombe, B. Loiseau,
                    and R. Vinh Mau,
                    Phys. Rev. Lett. {\bf 67}, 2423 (1991).
\bibitem{Tim84}     P.H. Timmers, W.A. van der Sanden, and
                    J.J. de Swart,
                    Phys. Rev. D {\bf 29}, 1928 (1984).
\bibitem{Liu90}     G.Q. Liu and F. Tabakin,
                    Phys. Rev. C {\bf 41}, 665 (1990).
\bibitem{Tim91c}    R. Timmermans, Ph.D. thesis, University of
                    Nijmegen, The Netherlands (1991).
\bibitem{Tim91b}    R. Timmermans, Th.A. Rijken, and
                    J.J. de Swart,
                    Phys. Rev. Lett. {\bf 67}, 1074 (1991).
\bibitem{Nag78}     M.M. Nagels, T.A. Rijken, and J.J. de Swart,
                    Phys. Rev. D {\bf 17}, 768 (1978).
\bibitem{Tim91a}    R. Timmermans, Th.A. Rijken, and
                    J.J. de Swart,
                    Phys. Lett. B {\bf 257}, 227 (1991).
\bibitem{Tim92a}    R. Timmermans, Th.A. Rijken, and
                    J.J. de Swart,
                    Phys. Rev. D {\bf 45}, 2288 (1992).
\bibitem{Par70}     M.H. Partovi and E.L. Lomon,
                    Phys. Rev. D {\bf 2}, 1999 (1970).
\bibitem{Erk74}     K. Erkelenz, Phys. Rep. {\bf C13}, 191 (1974).
\bibitem{Lip50}     B.A. Lippmann and J. Schwinger,
                    Phys. Rev. {\bf 79}, 469 (1950).
\bibitem{Log63}     A.A. Logunov and A.N. Tavkhelidze,
                    Il Nuovo Cimento {\bf 29}, 380 (1963).
\bibitem{Bla66}     R. Blankenbecler and R. Sugar,
                    Phys. Rev. {\bf 142}, 1051 (1966).
\bibitem{Kad68a}    V.G. Kadyshevsky, Nucl. Phys. {\bf B6}, 125 (1968).
\bibitem{Kad68b}    V.G. Kadyshevsky and M.D. Mateev,
                    Il Nuovo Cimento A {\bf 55}, 275 (1968).
\bibitem{Sal51}     E.E. Salpeter and H.A. Bethe,
                    Phys. Rev. {\bf 84}, 1232 (1951).
\bibitem{Gel51}     M. Gell-Mann and F. Low,
                    Phys. Rev. {\bf 84}, 350 (1951).
\bibitem{Nag77}     M.M. Nagels, T.A. Rijken, and J.J. de Swart,
                    Phys. Rev. D {\bf 15}, 2547 (1977).
\bibitem{Swa78}     J.J. de Swart and M.M. Nagels,
                    Fortschr. Phys. {\bf 26}, 215 (1978).
\bibitem{Ber87}     J.R. Bergervoet, Ph.D. thesis, University of
                    Nijmegen, (1987).
\bibitem{Wol52}     L. Wolfenstein and J. Ashkin,
                    Phys. Rev. {\bf 85}, 947 (1952).
\bibitem{Hos68}     N. Hoshizaki,
                    Prog. Theor. Phys. Suppl. {\bf 42}, 107 (1968).
\bibitem{Bys78}     J. Bystricky, F. Lehar, and P. Winternitz,
                    J. Phys. (Paris) {\bf 39}, 1 (1978).
\bibitem{LaF80}     P. LaFrance and P. Winternitz,
                    J. Phys. (Paris) {\bf 41}, 1391 (1980).
\bibitem{Bys84}     J. Bystricky, F. Lehar, and P. Winternitz,
                    J. Phys. (Paris) {\bf 45}, 207 (1984).
\bibitem{LaF92}     P. LaFrance, F. Lehar, B. Loiseau,
                    and P. Winternitz,
                    Helv. Phys. Acta {\bf 65}, 611 (1992).
\bibitem{Fes64}     H. Feshbach and E.L. Lomon,
                    Ann. Phys. (NY) {\bf 29}, 19 (1964).
\bibitem{Jaf79}     R.L. Jaffe and F.E. Low,
                    Phys. Rev. D {\bf 19}, 2105 (1979).
\bibitem{Jaf83}     R.L. Jaffe, in {\it Asymptotic Realms of Physics},
                    edited by A.H. Guth, K. Huang, and R.L. Jaffe
                    (The MIT Press, 1983), p. 100.
\bibitem{Bak86}     B.L.G. Bakker and P.J. Mulders,
                    Adv. Nucl. Phys. {\bf 17}, 1 (1986).
\bibitem{Swa85}     J.J. de Swart, J.R.M. Bergervoet, P.C.M. van Campen,
                    and W.A.M. van der Sanden, in {\it Advanced Methods in
                    the Evaluation of Nuclear Scattering Data}, Vol. 236
                    of {\it Lecture Notes in Physics}, edited by H.J.
                    Krappe and R. Lipperheide
                    (Springer-Verlag, Berlin, 1985), p. 179.
\bibitem{Fes58}     H. Feshbach, Ann. Phys. (NY) {\bf 5}, 357 (1958);
                    {\it ibid}. {\bf 19}, 287 (1962); {\bf 43}, 410 (1967).
\bibitem{Bre55}     G. Breit, Phys. Rev. {\bf 99}, 1581 (1955).
\bibitem{Aus83}     G.J.M. Austen and J.J. de Swart,
                    Phys. Rev. Lett. {\bf 50}, 2039 (1983).
\bibitem{Dur57}     L. Durand III, Phys. Rev. {\bf 108}, 1597 (1957).
\bibitem{Sto93}     V. Stoks, R. Timmermans, and J.J. de Swart,
                    Phys. Rev. C {\bf 47}, 512 (1993).
\bibitem{Mae89}     P.M.M. Maessen, Th.A. Rijken, and J.J. de Swart,
                    Phys. Rev. C {\bf 40}, 2226 (1989).
\bibitem{Swa89}     J.J. de Swart, T.A. Rijken, P.M. Maessen,
                    and R. Timmermans,
                    Il Nuovo Cimento {\bf 102A}, 203 (1989).
\bibitem{Hos60}     N. Hoshizaki, I. Lin, and
                    S. Machida, Prog. Theor. Phys.
                    (Kyoto) {\bf 24}, 480 (1960).
\bibitem{Hos62}     N. Hoshizaki, S. Otsuki, W. Watari,
                    and M. Yonezawa, Prog. Theor. Phys.
                    (Kyoto) {\bf 27}, 1199 (1962).
\bibitem{Bry64}     R.A. Bryan and B.L. Scott,
                    Phys. Rev. {\bf 135}, B434 (1964);
                    {\it ibid}. {\bf 177}, 1435 (1969).
\bibitem{Bin71}     J. Binstock and R.A. Bryan,
                    Phys. Rev. D {\bf 4}, 1341 (1971).
\bibitem{Bry72}     R.A. Bryan and A. Gersten,
                    Phys. Rev. D {\bf 6}, 341 (1972).
\bibitem{Pro73}     S.D. Protopopescu, M. Alston-Garnjost,
                    A. Barbaro-Galtieri, S.M. Flatt\'e,
                    J.H. Friedman, T.A. Lasinski, G.R. Lynch,
                    M.S. Rabin, and F.T. Solmitz,
                    Phys. Rev. D {\bf 7}, 1279 (1973).
\bibitem{Sve92b}    M. Svec, A. de Lesquen, and L. van Rossum,
                    Phys. Rev. D {\bf 46}, 949 (1992).
\bibitem{Sve92a}    M. Svec, A. de Lesquen, and L. van Rossum,
                    Phys. Rev. D {\bf 45}, 1518 (1992).
\bibitem{Jaf77}     R.L. Jaffe,
                    Phys. Rev. D {\bf 15}, 267 (1977).
\bibitem{Aer80}     A.T. Aerts, P.J. Mulders, and J.J. de Swart,
                    Phys. Rev. D {\bf 21}, 1370 (1980).
\bibitem{Sha84}     S.R. Sharpe, R.L. Jaffe, and
                    M.R. Pennington,
                    Phys. Rev. D {\bf 30}, 1013 (1984).
\bibitem{Eng78}     J.J. Engelen, M.J. Holwerda, E.W. Kittel,
                    H.G.J.M. Tiecke, J.S.M. Vergeest, B. Jongejans,
                    G.G.G. Massaro, H. Voorthuis, R.J. Hemingway,
                    S.O. Holmgren, M.J. Losty, S. Yamashita,
                    P. Grossmann, L. Lyons, and L. McDowell,
                    Nucl. Phys. {\bf B134}, 14 (1978).
\bibitem{Low75}     F.E. Low,
                    Phys. Rev. D {\bf 12}, 163 (1975).
\bibitem{Nus75}     S. Nussinov,
                    Phys. Rev. Lett. {\bf 34}, 1286 (1975).
\bibitem{Sim90}     Yu.A. Simonov,
                    Phys. Lett. B {\bf 249}, 514 (1990).
\bibitem{Dov78}     C.B. Dover and J.-M. Richard,
                    Phys. Rev. D {\bf 17}, 1770 (1978).
\bibitem{Dov79}     C.B. Dover and J.-M. Richard,
                    Ann. Phys. (NY) {\bf 121}, 70 (1979).
\bibitem{Buc79}     W. Buck, C.B. Dover, and J.-M. Richard,
                    Ann. Phys. (NY) {\bf 121}, 47 (1979).
\bibitem{Sch71}     G. Schierholz and S. Wagner,
                    Nucl. Phys. {\bf B32}, 306 (1971).
\bibitem{Bog74}     L.N. Bogdanova, O.D. Dalkarov, and I.S. Shapiro,
                    Ann. Phys. (NY) {\bf 84}, 261 (1974).
\bibitem{Sha78}     I.S. Shapiro,
                    Phys. Rep. {\bf 35C}, 129 (1978).
\bibitem{Mon80}     L. Montanet, G.C. Rossi, and G. Veneziano,
                    Phys. Rep. {\bf 63}, 153 (1980).
\bibitem{Car91}     J. Carbonell, O.D. Dalkarov, K.V. Protasov,
                    and I.S. Shapiro,
                    Nucl. Phys. {\bf A535}, 651 (1991).
\bibitem{Tay74}     J.R. Taylor,
                    Il Nuovo Cimento {\bf 23B}, 313 (1974).
\bibitem{Sem75}     M.D. Semon and J.R. Taylor,
                    Il Nuovo Cimento {\bf 26A}, 48 (1975).
\bibitem{Ger76}     A. Gersten,
                    Nucl. Phys. {\bf B103}, 465 (1976).
\bibitem{Sto90}     V.G.J. Stoks and J.J. de Swart,
                    Phys. Rev. C {\bf 42}, 1235 (1990).
\bibitem{Knu78}     L.D. Knutson and D. Chiang,
                    Phys. Rev. C {\bf 18}, 1958 (1978).
\bibitem{San99}     W.A. van der Sanden, unpublished.
\bibitem{Bla52}     J.M. Blatt and L.C. Biedenharn,
                    Phys. Rev. {\bf 86}, 399 (1952).
\bibitem{Sta57}     H.P. Stapp, T.J. Ypsilantis, and
                    N. Metropolis,
                    Phys. Rev. {\bf 105}, 302 (1957).
\bibitem{Bry81}     R.A. Bryan,
                    Phys. Rev. C {\bf 24}, 2659 (1981);
                    {\it ibid}. {\bf 30}, 305 (1984);
                    {\bf 39}, 783 (1989).
\bibitem{Spr82}     D.W.L. Sprung and M.W. Kermode,
                    Phys. Rev. C {\bf 26}, 1327 (1982).
\bibitem{Kla83}     S. Klarsfeld,
                    Phys. Lett. {\bf 126B}, 148 (1983).
\bibitem{Spr85}     D.W.L. Sprung,
                    Phys. Rev. C {\bf 32}, 699 (1985);
                    {\it ibid}. {\bf 35}, 869 (1987).
\bibitem{Arn66}     R.A. Arndt and M.H. MacGregor, in
                    {\it Methods in Computational Physics},
                    Vol. 6, edited by B. Alder, S. Fernbach, and
                    M. Rotenberg (Academic Press, 1966), p. 253.
%---- experimental data in the order as they appear in the table.
\bibitem{Ham80a}    R.P. Hamilton, T.P. Pun, R.D. Tripp,
                    H. Nicholson, D.M. Lazarus,
                    Phys. Rev. Lett. {\bf 44}, 1179 (1980).
\bibitem{Bru90}     W. Br\"uckner, B. Cujec, H. D\"obbeling,
                    K. Dworschak, F. G\"uttner, H. Kneis, S. Majewski,
                    M. Nomachi, S. Paul, B. Povh, R.D. Ransome,
                    T.-A. Shibata, M. Treichel, and Th. Walcher,
                    Zeit. Phys. {\bf A335}, 217 (1990).
\bibitem{Bru86a}    W. Br\"uckner, H. D\"obbeling, F. G\"uttner,
                    D. von Harrach, H. Kneis, S. Majewski, M. Nomachi,
                    S. Paul, B. Povh, R.D. Ransome, T.-A. Shibata,
                    M. Treichel, and Th. Walcher,
                    Phys. Lett. {\bf 166B}, 113 (1986).
\bibitem{Bru91a}    W. Br\"uckner, B. Cujec, H. D\"obbeling,
                    K. Dworschak, H. Kneis, S. Majewski, M. Nomachi,
                    S. Paul, B. Povh, R.D. Ransome, T.-A. Shibata,
                    M. Treichel, and Th. Walcher,
                    Zeit. Phys. {\bf A339}, 367 (1991).
\bibitem{Bru86b}    W. Br\"uckner, H. D\"obbeling, F. G\"uttner,
                    D. von Harrach, H. Kneis, S. Majewski,
                    M. Nomachi, S. Paul, B. Povh, R.D. Ransome,
                    T.-A. Shibata, M. Treichel, and Th. Walcher,
                    Phys. Lett. {\bf 169B}, 302 (1986).
\bibitem{Spe70}     D. Spencer and D.N. Edwards,
                    Nucl. Phys. {\bf B19}, 501 (1970).
\bibitem{Bru87}     W. Br\"uckner, B. Cujec, H. D\"obbeling,
                    K. Dworschak, F. G\"uttner, H. Kneis, S. Majewski,
                    M. Nomachi, S. Paul, B. Povh, R.D. Ransome,
                    T.-A. Shibata, M. Treichel, and Th. Walcher,
                    Phys. Lett. B {\bf 197}, 463 (1987);
                    {\it ibid}. {\bf 199}, 596(E) (1987).
\bibitem{Bug87}     D.V. Bugg, J. Hall, A.S. Clough, R.L. Shypit,
                    K. Bos, J.C. Kluyver, R.A. Kunne, L. Linssen,
                    R. Birsa, F. Bradamante,
                    S. Dalla Torre-Colautti, A. Martin,
                    A. Penzo, P. Schiavon, A. Villari,
                    E. Heer, C. LeLuc, Y. Onel, and D. Rapin,
                    Phys. Lett. B {\bf 194}, 563 (1987).
\bibitem{Lin87}     L. Linssen, C.I. Beard, R. Birsa, K. Bos,
                    F. Bradamante, D.V. Bugg, A.S. Clough,
                    S. Dalla Torre-Colautti, M. Giorgi, J.R. Hall,
                    J.C. Kluyver, R.A. Kunne, C. Lechanoine-LeLuc,
                    A. Martin, Y. Onel, A. Penzo, D. Rapin,
                    P. Schiavon, R.L. Shypit, and A. Villari,
                    Nucl. Phys. {\bf A469}, 726 (1987).
\bibitem{Als75}     M. Alston-Garnjost, R. Kenney, D. Pollard,
                    R. Ross, R. Tripp, and H. Nicholson,
                    Phys. Rev. Lett. {\bf 35}, 1685 (1975).
\bibitem{Sum82}     T. Sumiyoshi, J. Chiba, T. Fujii, H. Iwasaki,
                    T. Kageyama, S. Kuribayashi, K. Nakamura,
                    T. Takeda, H. Ikeda, and Y. Takada,
                    Phys. Rev. Lett. {\bf 49}, 628 (1982).
\bibitem{Con68}     B. Conforto, G. Fidecaro, H. Steiner,
                    R. Bizzarri, P. Guidoni, F. Marcelja,
                    G. Brautti, E. Castelli, M. Ceschia, and M. Sessa,
                    Il Nuovo Cimento A {\bf 54}, 441 (1968).
\bibitem{Ham80b}    R.P. Hamilton, T.P. Pun, R.D. Tripp,
                    D.M. Lazarus, and H. Nicholson,
                    Phys. Rev. Lett. {\bf 44}, 1182 (1980).
\bibitem{Cre83}     M. Cresti, L. Peruzzo, and G. Sartori,
                    Phys. Lett. {\bf 132B}, 209 (1983).
\bibitem{Sak82}     S. Sakamoto, T. Hashimoto, F. Sai,
                    and S.S. Yamamoto,
                    Nucl. Phys. {\bf B195}, 1 (1982).
\bibitem{Clo84}     A.S. Clough, C.I. Beard, D.V. Bugg,
                    J.A. Edgington, J. Hall, K. Bos, J.C. Kluyver,
                    R.A. Kunne, L. Linssen, R. Birsa, F. Bradamante,
                    S. Dalla Torre-Colautti, M. Giorgi, A. Martin,
                    A. Penzo, P. Schiavon, A. Villari,
                    S. Degli-Agosti, E. Heer, R. Hess,
                    C. Lechanoine-LeLuc, Y. Onel, and D. Rapin,
                    Phys. Lett. {\bf 146B}, 299 (1984).
\bibitem{Kag87}     T. Kageyama, T. Fujii, K. Nakamura, F. Sai,
                    S. Sakamoto, S. Sato, T. Takahashi, T. Tanimori,
                    S.S. Yamamoto, and Y. Takada,
                    Phys. Rev. D {\bf 35}, 2655 (1987).
\bibitem{Nak84b}    K. Nakamura, T. Fujii, T. Kageyama, F. Sai,
                    S. Sakamoto, S. Sato, T. Takahashi,
                    T. Tanimori, and S.S. Yamamoto,
                    Phys. Rev. Lett. {\bf 53}, 885 (1984).
\bibitem{Als79}     M. Alston-Garnjost, R.P. Hamilton,
                    R.W. Kenney, D.L. Pollard, R.D. Tripp,
                    H. Nicholson, and D.M. Lazarus,
                    Phys. Rev. Lett. {\bf 43}, 1901 (1979).
\bibitem{Iwa81}     H. Iwasaki, H. Aihara, J. Chiba, H. Fujii,
                    T. Fujii, T. Kamae, K. Nakamura, T. Sumiyoshi,
                    Y. Takada, T. Takeda, M. Yamauchi, and H. Fukuma,
                    Phys. Lett. {\bf 103B}, 247 (1981).
\bibitem{Biz68}     R. Bizzarri, B. Conforto, G.C. Gialanella,
                    P. Guidoni, F. Marcelja, E. Castelli,
                    M. Ceschia, and M. Sessa,
                    Il Nuovo Cimento A {\bf 54}, 456 (1968).
\bibitem{Per91}     F. Perrot-Kunne, R. Bertini, M. Costa, H. Catz,
                    A. Chaumeaux, J.-C. Faivre, E. Vercellin,
                    J. Arvieux, J. Yonnet, B. van den Brandt,
                    D.R. Gill, J.A. Konter, S. Mango, G.D. Wait,
                    E. Boschitz, W. Gyles, W. List, C. Otterman,
                    R. Tacik, and M. Wessler,
                    Phys. Lett. B {\bf 261}, 188 (1991).
\bibitem{Tsu83}     T. Tsuboyama, Y. Kubota, F. Sai, S. Sakamoto,
                    and S.S. Yamamoto,
                    Phys. Rev. D {\bf 28}, 2135 (1983).
\bibitem{Kun88}     R.A. Kunne, C.I. Beard, R. Birsa, K. Bos,
                    F. Bradamante, D.V. Bugg, A.S. Clough,
                    S. Dalla Torre-Colautti, S. Degli-Agosti,
                    J.A. Edgington, J.R. Hall, E. Heer, R. Hess,
                    J.C. Kluyver, C. Lechanoine-LeLuc, L. Linssen,
                    A. Martin, T.O. Niinikoski, Y. Onel, A. Penzo,
                    D. Rapin, J.M. Rieubland, A. Rijllart,
                    P. Schiavon, R.L. Shypit, F. Tessarotto,
                    A. Villari, and P. Wells,
                    Phys. Lett. B {\bf 206}, 557 (1988).
\bibitem{Kun89}     R.A. Kunne, C.I. Beard, R. Birsa, K. Bos,
                    F. Bradamante, D.V. Bugg, A.S. Clough,
                    S. Dalla Torre-Colautti,
                    J.A. Edgington, J.R. Hall, E. Heer, R. Hess,
                    J.C. Kluyver, C. Lechanoine-LeLuc, L. Linssen,
                    A. Martin, T.O. Niinikoski, Y. Onel, A. Penzo,
                    D. Rapin, J.M. Rieubland, A. Rijllart,
                    P. Schiavon, R.L. Shypit, F. Tessarotto,
                    A. Villari, and P. Wells,
                    Nucl. Phys. {\bf B323}, 1 (1989).
\bibitem{Bir91}     R. Birsa, F. Bradamante,
                    A. Bressan, S. Dalla Torre-Colautti,
                    M. Giorgi, M. Lamanna, A. Martin, A. Penzo,
                    P. Schiavon, F. Tessarotto, M.P. Macciotta,
                    A. Masoni, G. Puddu, S. Serci, T. Niinikoski,
                    A. Rijllart, A. Ahmidouch, E. Heer, R. Hess,
                    C. Lechanoine-LeLuc, C. Mascarini, D. Rapin,
                    J. Arvieux, R. Bertini, H. Catz, J.C. Faivre,
                    R.A. Kunne, F. Perrot-Kunne, M. Agnello,
                    F. Iazzi, B. Minetti, T. Bressani, E. Chiavassa,
                    N. De Marco, A. Musso, and A. Piccotti,
                    Phys. Lett. B {\bf 273}, 533 (1991).
\bibitem{Sch89}     P. Schiavon, R. Birsa, K. Bos, F. Bradamante,
                    A.S. Clough, S. Dalla Torre-Colautti, J.R. Hall,
                    E. Heer, R. Hess, J.C. Kluyver, R.A. Kunne,
                    C. Lechanoine-LeLuc, L. Linssen,
                    A. Martin, Y. Onel, A. Penzo, D. Rapin,
                    R.L. Shypit, F. Tessarotto, and A. Villari,
                    Nucl. Phys. {\bf A505}, 595 (1989).
\bibitem{Bir90}     R. Birsa, F. Bradamante, S. Dalla Torre-Colautti,
                    M. Giorgi, M. Lamanna, A. Martin, A. Penzo,
                    P. Schiavon, F. Tessarotto, M.P. Macciotta,
                    A. Masoni, G. Puddu, S. Serci, T. Niinikoski,
                    A. Rijllart, A. Ahmidouch, E. Heer, R. Hess,
                    R.A. Kunne, C. Lechanoine-LeLuc, C. Mascarini,
                    D. Rapin, J. Arvieux, R. Bertini, H. Catz,
                    J.C. Faivre, F. Perrot-Kunne, M. Agnello,
                    F. Iazzi, B. Minetti, T. Bressani, E. Chiavassa,
                    N. De Marco, A. Musso, and A. Piccotti,
                    Phys. Lett. B {\bf 246}, 267 (1990).
\bibitem{Ohs73}     T. Ohsugi, M. Fujisaki, S. Kaneko, Y. Murata,
                    K. Okamura, H. Kohno, M. Fukawa, R. Hamatsu,
                    T. Hirose, S. Kitamura, T. Mamiya, T. Yamagata,
                    T. Emura, I. Kita, and K. Takahashi,
                    Il Nuovo Cimento {\bf 17A}, 456 (1973).
\bibitem{Eis76}     E. Eisenhandler, W.R. Gibson, C. Hojvat,
                    P.I.P. Kalmus, L.C.Y. Lee, T.W. Pritchard,
                    E.C. Usher, D.T. Williams, M. Harrison,
                    W.H. Range, M.A.R. Kemp, A.D. Rush, J.N. Woulds,
                    G.T.J. Arnison, A. Astbury, D.P. Jones, and
                    A.S.L. Parsons,
                    Nucl. Phys. {\bf B113}, 1 (1976).
\bibitem{Koh72}     H. Kohno, S. Kaneko, Y. Murata, T. Ohsugi,
                    K. Okamura, M. Fukawa, R. Hamatsu, T. Hirose,
                    T. Mamiya, T. Yamagata, T. Emura, I. Kita,
                    and K. Takahashi,
                    Nucl. Phys. {\bf B41}, 485 (1972).
\bibitem{Ber89}     R. Bertini, M. Costa, F. Perrot, H. Catz,
                    A. Chaumeaux, J.Cl. Faivre, E. Vercellin,
                    J. Arvieux, J. Yonnet, B. van den Brandt,
                    J.A. Konter, D.R. Gill, S. Mango, G.D. Wait,
                    E. Boschitz, W. Gyles, W. List, C. Otterman,
                    R. Tacik, M. Wessler, E. Descroix,
                    J.Y. Grossiord, and A. Guichard,
                    Phys. Lett. B {\bf 228}, 531 (1989).
\bibitem{Kim82}     M. Kimura, M. Takanaka, R. Hamatsu,
                    Y. Hattori, T. Hirose, S. Kitamura,
                    T. Yamagata, T. Emura, I. Kita, K. Takahashi,
                    H. Kohno, and S. Matsumoto,
                    Il Nuovo Cimento {\bf 71A}, 438 (1982).
\bibitem{Alb72}     M.G. Albrow, S. Andersson/Almehed,
                    B. Bo\v{s}njakovi\'c, C. Daum, F.C. Ern\'e
                    Y. Kimura, J.P. Lagnaux, J.C. Sens, and F. Udo,
                    Nucl. Phys. {\bf B37}, 349 (1972).
%---- other data, e.g. data rejected from the start.
\bibitem{Car74}     A.S. Carroll, I-H. Chiang, T.F. Kycia, K.K. Li,
                    P.O. Mazur, D.N. Michael, P. Mockett, D.C. Rahm,
                    and R. Rubinstein,
                    Phys. Rev. Lett. {\bf 32}, 247 (1974).
\bibitem{Cha76}     V. Chaloupka, H. Dreverman, F. Marzano,
                    L. Montanet, P. Schmid, J.R. Fry, H. Rohringer,
                    S. Simopoulou, J. Hanton, F. Grard, V.P. Henri,
                    H. Johnstad, J.M. Lesceux, J.S. Skura, A. Bettini,
                    M. Cresti, L. Peruzzo, P. Rossi, R. Bizzarri,
                    M. Iori, E. Castelli, C. Omero, and P. Poropat,
                    Phys. Lett. {\bf 61B}, 487 (1976).
\bibitem{Sak79}     S. Sakamoto, T. Hashimoto, F. Sai,
                    and S.S. Yamamoto,
                    Nucl. Phys. {\bf B158}, 410 (1979).
\bibitem{Kam80}     T. Kamae, H. Aihara, J. Chiba, H. Fujii,
                    T. Fujii, H. Iwasaki, K. Nakamura, T. Sumiyoshi,
                    Y. Takada, T. Takeda, M. Yamauchi, H. Fukuma,
                    and T. Takeshita,
                    Phys. Rev. Lett. {\bf 44}, 1439 (1980).
\bibitem{Nak84a}    K. Nakamura, H. Aihara, J. Chiba, H. Fujii,
                    T. Fujii, H. Iwasaki, T. Kamae, T. Sumiyoshi,
                    Y. Takada, T. Takeda, M. Yamauchi, H. Fukuma,
                    and T. Takeshita,
                    Phys. Rev. D {\bf 29}, 349 (1984).
\bibitem{Tim85}     P.H. Timmers, W.A. van der Sanden, and
                    J.J. de Swart,
                    Phys. Rev. D {\bf 31}, 99 (1985).
\bibitem{Bru77}     W. Br\"uckner, B. Granz, D. Ingham, K. Kilian,
                    U. Lynen, J. Niewisch, B. Pietrzyk, B. Povh,
                    H.G. Ritter, and H. Schr\"oder,
                    Phys. Lett. {\bf 67B}, 222 (1977).
\bibitem{Jas81}     E. Jastrzembski, N. Haik, W.K. McFarlane,
                    M.A. Mandelkern, D.C. Schultz, C. Amsler,
                    C.C. Herrmann, and D.M. Wolfe,
                    Phys. Rev. D {\bf 23}, R2784 (1981).
\bibitem{Low81}     D.I. Lowenstein, D.C. Peaslee, C. Bromberg,
                    R.A. Lewis, R. Miller, B.Y. Oh, T. Potter,
                    G.A. Smith, J. Whitmore, T. Brando,
                    T.E. Kalogeropoulos, G. Onengut, C. Petridou,
                    M. Singer, and G.S. Tzanakos,
                    Phys. Rev. D {\bf 23}, R2788 (1981).
\bibitem{Sai83}     F. Sai, S. Sakamoto, and S.S. Yamamoto,
                    Nucl. Phys. {\bf B213}, 371 (1983).
\bibitem{Bra85}     T. Brando, I. Daftari, A. Deguzman,
                    T.E. Kalogeropoulos, R.A. Lewis, D.I. Lowenstein,
                    R.J. Miller, B.Y. Oh, D.C. Peaslee, C. Petridou,
                    M. Singer, G.A. Smith, G.S. Tzanakos,
                    R. Venugopal, R.D. von Lintig, and J. Whitmore,
                    Phys. Lett. {\bf 158B}, 505 (1985).
\bibitem{Fra87}     J. Franklin,
                    Phys. Lett. B {\bf 184}, 111 (1987).
\bibitem{Cou77}     M. Coupland, E. Eisenhandler, W.R. Gibson,
                    P.I.P. Kalmus, and A. Astbury,
                    Phys. Lett. {\bf 71B}, 460 (1977).
\bibitem{Kas78}     H. Kaseno,
                    Il Nuovo Cimento {\bf 43A}, 119 (1978).
\bibitem{Ash85}     V. Ashford, M.E. Sainio, M. Sakitt, J. Skelly,
                    R. Debbe, W. Fickinger, R. Marino,
                    and D.K. Robinson,
                    Phys. Rev. Lett. {\bf 54}, 518 (1985).
\bibitem{Bru85}     W. Br\"uckner, H. D\"obbeling, F. G\"uttner,
                    D. von Harrach, H. Kneis, S. Majewski, M. Nomachi,
                    S. Paul, B. Povh, R.D. Ransome, T.-A. Shibata,
                    M. Treichel, and Th. Walcher,
                    Phys. Lett. {\bf 158B}, 180 (1985).
\bibitem{Bru91b}    W. Br\"uckner, B. Cujec, H. D\"obbeling,
                    K. Dworschak, H. Kneis, S. Majewski, M. Nomachi,
                    S. Paul, B. Povh, R.D. Ransome, T.-A. Shibata,
                    M. Treichel, and Th. Walcher,
                    Zeit. Phys. {\bf A339}, 379 (1991).
\bibitem{Kun91}     R.A. Kunne, C.I. Beard, R. Birsa, K. Bos,
                    F. Bradamante, D.V. Bugg, A.S. Clough,
                    S. Dalla Torre-Colautti, J.A. Edgington,
                    M. Giorgi, J.R. Hall, E. Heer, R. Hess,
                    J.C. Kluyver, C. Lechanoine-LeLuc, L. Linssen,
                    A. Martin, T.O. Niinikoski, Y. Onel, A. Penzo,
                    D. Rapin, J.M. Rieubland, A. Rijllart,
                    P. Schiavon, R.L. Shypit, F. Tessarotto,
                    and A. Villari,
                    Phys. Lett. B {\bf 261}, 191 (1991).
\bibitem{Bog76}     M. Bogdanski, T. Emura, S.N. Ganguli, A. Gurtu,
                    S. Hamada, R. Hamatsu, E. Jeannet, I. Kita,
                    S. Kitamura, J. Kishiro, H. Kohno, M. Komatsu,
                    P.K. Malhorta, S. Matsumoto, U. Mehtani,
                    L. Montanet, R. Raghavan, A. Subramanian,
                    K. Takahashi, and T. Yamagata,
                    Phys. Lett. {\bf 62B}, 117 (1976).
\bibitem{Ban85}     S. Banerjee, S.N. Ganguli, A. Gurtu,
                    P.K. Malhorta, R. Raghavan, A. Subramanian,
                    K. Sudhakar, M.M. Agarwal, J.M. Kohli, J.P. Lamba,
                    I.S. Mittra, J.B. Singh, P.M. Sood, Dev Anand,
                    P.V.K.S. Baba, G.L. Kaul, Y. Prakash, N.K. Rao,
                    G. Singh, R. Hamatsu, T. Hirose, S. Kitamura,
                    and T. Yamagata,
                    Zeit. Phys. {\bf C28}, 163 (1985).
\bibitem{Mac92}     M.P. Macciotta, A. Masoni, G. Puddu, S. Serci,
                    A. Ahmidouch, E. Heer, C. Mascarini, D. Rapin,
                    J. Arvieux, R. Bertini, J.C. Faivre, R.A. Kunne,
                    R. Birsa, F. Bradamante (Spokesman), A. Bressan,
                    S. Dalla Torre-Colautti, M. Giorgi, M. Lamanna,
                    A. Martin, A. Penzo, P. Schiavon,
                    F. Tessarotto, A.M. Zanetti,
                    E. Chiavassa, N. De Marco, A. Musso, and
                    A. Piccotti, (The PS199 Collaboration),
                    {\it Proposal to the CERN SPLSC. Measurement of
                    the \mbox{$\overline{p}p \rightarrow \overline{n}n$}
                    charge-exchange differential cross section},
                    CERN/SPSLC 92-17 (1992).
\bibitem{Lea76}     E. Leader, Phys. Lett. {\bf 60B}, 290 (1976).
\bibitem{Klo91}     R.A.M. Klomp, V.G.J. Stoks, and J.J. de Swart,
                    Phys. Rev. C {\bf 44}, R1258 (1991).
\bibitem{Arn90}     R.A. Arndt, Z. Li, L.D. Roper, and R.L. Workman,
                    Phys. Rev. Lett. {\bf 65}, 157 (1990).
\end{references}
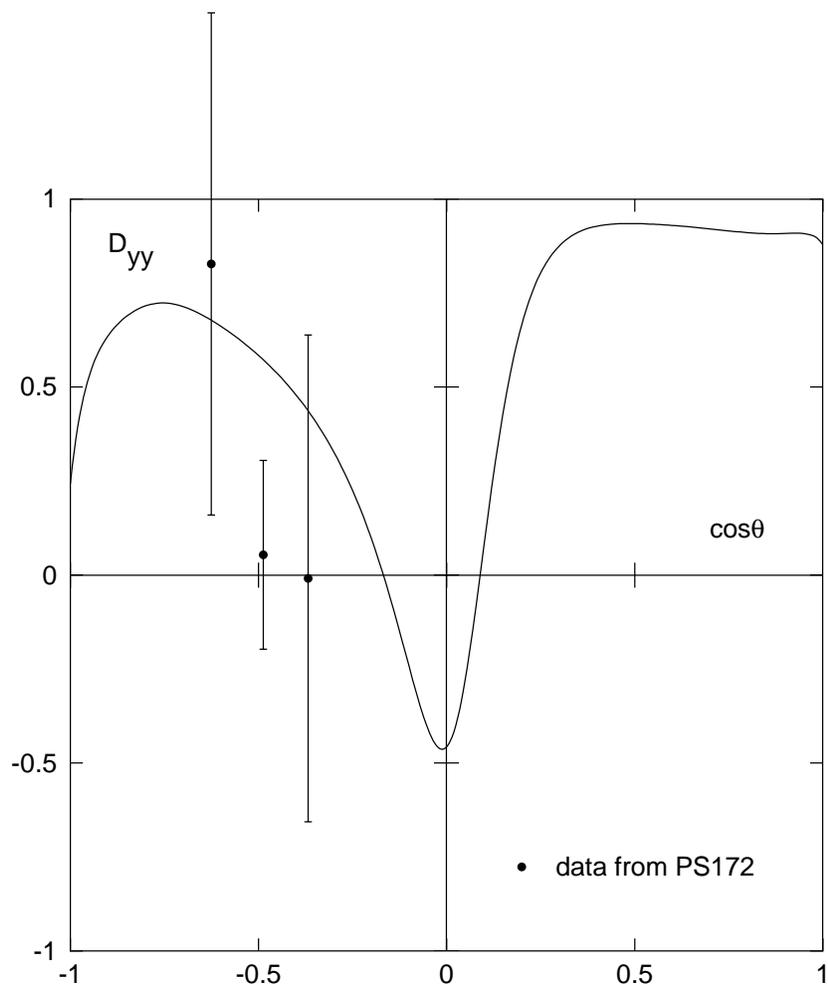
\end{document}